\documentclass[a4paper,fleqn]{cas-sc}

\usepackage{pgfplots} 
\usepackage{longtable} 
\usepackage[online]{threeparttablex} 
\usepackage{array} 
\usepackage{amsmath}
\usepackage{lineno}  
\usepackage{etoolbox}

\pretocmd\equation{\linenomath}{}{}
\apptocmd\endequation{\endlinenomath}{}{}

\usepackage{rotating}
\usepackage{caption}
\usepackage{subcaption}

\usepackage{hyperref}
\hypersetup{colorlinks=true,linkcolor=blue, citecolor=blue,  linktocpage}

\newcommand\inputpgf[2]{{
		\let\includegraphicsWithoutPath\includegraphics
		\renewcommand{\includegraphics}[2][]{\includegraphicsWithoutPath[##1]{#1/##2}}
		\input{#1/#2}
}}
\pgfplotsset{compat=1.17} 

\usepackage{adjustbox} 
\usepackage{printlen} 

\usepackage{supertabular} 
\usepackage{booktabs} 

\usepackage{hyperref,xcolor} 

\usepackage[authoryear]{natbib}

\def\tsc#1{\csdef{#1}{\textsc{\lowercase{#1}}\xspace}}
\tsc{WGM}
\tsc{QE}

\begin{document}
\let\WriteBookmarks\relax
\def\floatpagepagefraction{1}
\def\textpagefraction{.001}

\shorttitle{}    

\shortauthors{}  

\title [mode = title]{New Insights into Erg Chech 002 Analogues in the Main Belt from Near-Infrared Spectroscopy.}  



%

\author[1,3]{M. Galinier}[orcid=0000-0001-7920-0133]



\ead{marjorie.galinier@inaf.it}



\affiliation[1]{organization={Istituto di Astrofisica e Planetologia Spaziali, Istituto Nazionale di Astrofisica},
            addressline={via del Fosso del
            	Cavaliere, 100}, 
            city={Rome},
            postcode={00133},
            country={Italy}}

\author[2]{C. Avdellidou}


\ead{ca337@leicester.ac.uk}



\affiliation[2]{organization={School of Physics and Astronomy, University of Leicester},
            addressline={University Road}, 
            city={Leicester},
            postcode={LE1 7RH}, 
            country={UK}}
            
\author[3,2]{M. Delbo}


\ead{delbo@oca.eu}



\affiliation[3]{organization={Universit\'e C\^ote d'Azur, CNRS--Lagrange, Observatoire de la C\^ote d'Azur},
	addressline={96 Boulevard de l'Observatoire}, 
	city={Nice},
	postcode={06304}, 
	country={France}}

\author[3,4]{L. Galluccio}


\ead{laurent.galluccio@oca.eu}



\affiliation[4]{organization={INAF – Osservatorio Astronomico di Roma.},
	addressline={Via Frascati 33}, 
	city={Rome},
	postcode={I-00078 Monte Porzio Catone}, 
	country={Italy}}

\cortext[1]{Corresponding author}



\begin{abstract}
Erg Chech 002 is an andesitic meteorite that formed early in the Solar System's history, and that is thought to have been formed in the primitive crust of an early accreted and differentiated planetesimal.  It shows unique spectral features, and no known asteroid spectral type was initially found to match with its compositional type. In the literature, asteroids (10537) 1991~RY16, (7472) Kumakiri and (14390) 1990~QP10 were found to show peculiar spectra, and were not classified in any known existing spectral class. These objects were hypothesised as well to be fragments of differentiated planetesimals. In a previous study, the Gaia Data Release 3 dataset of visible reflectance spectra of Solar System Objects was exploited to search for potential analogues of Erg Chech 002 in the main belt. As a result, 142 asteroids were found to potentially match this meteorite. In this work, we present NASA's IRTF near-infrared observations of 20 main belt asteroids found as potential analogues of Erg Chech 002. We classified these asteroids based on their visible and near-infrared spectra, then studied and compared their diagnostic spectral features (band centres and band area ratios) with those of the laboratory spectra of Erg Chech 002. We classified 16 of the 20 observed asteroids as V-types, one as S-complex and one as A-type, and conclude that none of the observed objects match with Erg Chech 002. In addition, we show that asteroids (10537) 1991~RY16 and (14390) 1990~QP10 are good spectral matches to Erg Chech 002 based on the study of their diagnostic spectral features. Asteroid (7472) Kumakiri could match a more olivine-rich EC~002-like material, and cannot be completely ruled out as a match of the meteorite. This possible link between the meteorite and these asteroids is consistent with the theories regarding their formation, and these asteroids could be part of a new spectral class of andesitic bodies in the main belt. \nocite{*}
\end{abstract}


\maketitle


\section{Introduction}

Asteroids are the remnants of planetesimals, bodies from some tens to a few hundreds of kilometres in size that formed in the first million years of our Solar System's history from the accretion of dust in the protoplanetary disk, and that are deemed as the `building blocks' of planets \citep{Morbidelli2009_bigast,henke2012,Klahr_Schreiber_2015,delbo2017,trieloff2022}. Due to the heat generated by the abundance of $^{26}$Al in forming planetesimals, some underwent complete or partial differentiation, with their interiors melting and organising into layers of different densities and compositions: an iron-rich core, an olivine-rich mantle and an igneous-crust \citep[see for example][and references therein]{McConnell1967,mcsween2002,Bizzarro_2005,Elkins-Tanton2011,Neumann2012diff}. These planetesimals evolved little from a chemical and thermal point of view after their cooling, and the asteroids populating the main belt today are collisional by-products of this original population of bodies \citep[see for example][]{bottke2005,Morbidelli2009_bigast,delbo2017,delbo2019,klahr2020,Ferrone2023}.

When an asteroid (the `parent body') undergoes a catastrophic collision, it is partially or completely disrupted, resulting in the formation of a family of fragments that share similar orbital elements and a composition reflecting the original composition of the parent body \citep{Ivezic2002,parker2008,Cellino2009}. The action of non-gravitational forces can lead some of these fragments to drift towards resonances in the main belt and to be subsequently delivered to Earth as meteorites \citep{greenwood2020}.


The classification of meteorites is based on the study of their mineralogical and chemical composition, and the total meteorite collection today (as of February 20, 2026) counts more than 86~000 meteorites divided into more than 40 different groups. Among differentiated meteorites, achondrites represent about 5\% of the total meteorite collection, and ungrouped achondrites only about 0.2\% (source: the Meteoritical Bulletin Database, February 20, 2026). Among the many groupings of differentiated meteorites, andesites belong to the ungrouped achondrites class due to their unique composition and petrology \citep{Day2009, barrat2021}. The meteorite Erg Chech 002 (hereafter EC~002) belongs to this class and has been designated by \cite{barrat2021} as `the oldest andesite of the Solar System', with an estimated age of about $\sim$$4,565$ Myr that has been inferred from diverse dating techniques \citep{barrat2021,Fang2022,Reger2023,Anand2022,Zhu2022,krestianinov_u-pb_2022,Connelly2023,Yang2025}. It likely formed from the partial melting of a chondritic source, as deduced from its petrology and composition \citep{Yamaguchi2021,Nicklas2021,CollinetGrove2020,Chaussidon2021}. More specifically, EC~002 is thought to have formed in a magmatic intrusion in the primitive igneous crust of a non-carbonaceous planetesimal that had undergone early accretion and differentiation \citep{Yamaguchi2021,Nicklas2021,CollinetGrove2020,Chaussidon2021,barrat2021,Beros2024}. The differentiation was likely partial, preserving enough chondritic material to produce EC~002 \citep{barrat2021,Anand2022,Neumann2023,Sturtz2022}. Another analysis based on the study of EC~002's xenocrysts suggests that this meteorite could have formed in a magma ocean, that has been altered by some mixing with meteoritic material from both the inner and outer Solar System \citep{Jin2024}. Evidence of rapid cooling and of shock deformation of the meteorite suggests that a violent event following the formation of EC~002, such as a hypervelocity impact, could have ejected EC~002 from the original parent body \citep{Reger2023,Neumann2023,Beros2024}. The size of the parent body is still under dispute: using a thermal model, \cite{Sturtz2022} deduced that its crust must have been a few km-thick, for a total radius between 70 and 130~km ; while \cite{Neumann2023} found a parent body of 20-30~km using thermo-chronological data.

\cite{barrat2021} suggest that the parent body of EC~002 was likely not unique, and bodies with andesitic crusts could have been common  $\sim$$4,565$ Myr ago. Therefore, finding the parent body of meteorites such as EC~002 in the main belt asteroid population would allow to study fragments of the earliest accreted and differentiated planetesimals, and would allow to further constrain the formation of the Solar System.

To search for potential analogues of EC~002 in the main belt asteroid population, \cite{barrat2021} acquired laboratory spectra of a powder and three slab samples of EC~002. These spectra show the presence of two strong absorption bands centred around 950 nm (Band I) and 2,000 nm (Band II), which have been associated with calcium-rich pyroxene; and the presence of a small band centred around 650 nm, the origin of which is not discussed in \cite{barrat2021}.

Establishing links between meteorites and asteroids makes it possible to study the asteroid population with greater precision and to place constraints on the formation and evolution of the Solar System, since meteorites provide a link between mineralogy and spectral properties \citep{Lucas2019}. To link asteroids with meteorites, classical techniques rely on the comparison of the bodies' spectral characteristics. As an example, V-type asteroid (4) Vesta and members of its dynamical family have been successfully linked to the group of Howardite, Eucrite and Diogenite meteorites (HEDs) \citep{McCord1970,russell2012}. The petrology of eucrites is consistent with a production in the superior crust of Vesta, while diogenites are understood to come from deeper crustal or upper mantle parts of the asteroid, and Howardites show an intermediate petrology \citep{McSween2013}. This genetic link along with the discovery of impact craters on (4) Vesta shed light on the basaltic composition of the Vesta family members, and on the family's formation and evolution \citep[see e.g.][and references therein]{Mansour2020}.




Establishing links between meteorites and asteroids is made difficult by various phenomena that affect the spectral characteristics and slope of asteroids, such as space weathering \citep{Hapke2001, PietersNoble2016}. Indeed, asteroids in space are subject to impacts from meteoroids and micrometeoroids, interaction with solar wind particles, and bombardment by cosmic rays \citep{clark2002}. These phenomena are known to alter the reflectance spectrum of an asteroid, depending on its composition. For silicate materials, it induces a darkening of an object's albedo, a reddening of its spectral slope, and a reduction in the depth of its silicate bands \citep{AdamsMcCord1971,brunetto2015}. Space weathering must be taken into account in order to link meteorites to asteroids, as meteorites' spectra are not affected by it. However, determining the corrections to be applied to an asteroid's spectrum to account for these effects is not trivial, as they appear to differ depending on the mineralogical composition of the objects \citep{Gaffey2011,Marchi2010,Fulvio2012,brunetto2015,Zhang2022}.

\cite{barrat2021} applied a space-weathering model adapted to silicate-rich asteroids \citep{Hapke2001} to the laboratory spectra of EC~002, in order to compare them with astronomical spectra of asteroids with strong pyroxene signatures. They used spectrophotometric data from the Sloan Digital Sky Survey (SDSS), and visible and near-infrared (VISNIR) astronomical spectra of asteroids available at the time in the literature. However, they did not find any satisfactory match of the meteorite in the asteroid population. \cite{barrat2021} therefore concluded that the original population of planetesimals and their fragments with compositions similar to EC~002 must have disappeared. They hypothesize that these objects have either been accreted by other planetesimals or destroyed. 

\cite{galinier2023} used the Gaia Data Release 3 (DR3) dataset of more than 60\,000 asteroid reflectance spectra spanning the visible (VIS) wavelength range \citep{galluccio2022} to search for potential analogues of EC~002 in the main belt. They studied the spectra of the four laboratory samples and three modelled space-weathered spectra of EC~002 respectively acquired and produced by \cite{barrat2021}. Using several spectral comparison techniques, they identified 142 asteroids with a Gaia DR3 spectrum indicating a potential andesitic composition: 0.08\% of the Gaia DR3 dataset potentially matching the laboratory samples of the meteorite, and 0.15\% matching its modelled space-weathered spectra. However, the search for potential analogues of EC~002 using the Gaia DR3 dataset is challenging because of some issues affecting the average reflectance spectra, such as a `fake band' that can appear around 650 nm related to the merging of the spectra acquired independently by the Blue and Red Spectrophotometers (BP and RP) onboard Gaia \citep{galluccio2022,galinier2023}. Unfortunately, when present, this artificial band is located where a real band is found in the spectrum of EC~002, which can cause a basaltic asteroid to appear as a potential analogue of the meteorite when considering only the VIS wavelength range. Therefore, the potential matches found in \cite{galinier2023} remained to be confirmed by observations in the near-infrared (NIR) wavelength range, as this range contains diagnostic features representative of specific mineralogies and allowing to distinguish basaltic from andesitic bodies, such as the depth of the bands around 1 and 2 $\mu$m and the band area ratio \citep {Gaffey2011}.

In the literature, some asteroids have been found to show peculiar spectra in the NIR wavelength range, and have been hypothesized to show a different mineralogy than the ones of known spectral types. It is the case of asteroids (10537) 1991~RY16 \citep{moskovitz2008,moskovitz2008_10537,leith2017,Hardersen2018}, (7472) Kumakiri \citep{Duffard2009,Hicks2014,leith2017} and (14390) 1990~QP10 \citep{leith2017,Hardersen2018,Migliorini2018,Matlovic2020}. These objects show spectra between those of V-, R- or O-types, with a band around 1 $\mu$m broader and shifted towards longer wavelengths compared to basaltic V-type asteroids, and a shallower 2 $\mu$m band \citep{moskovitz2008_10537,leith2017,Hardersen2018,Migliorini2018,Matlovic2020}. Moreover, (10537) 1991~RY16 and (7472) Kumakiri show the presence of a small band centred around 650 nm \citep{moskovitz2008_10537,Duffard2009,leith2017} that is not observed with such intensity in basaltic asteroids \citep{Duffard2009} and that is reminiscent of the band observable in the spectrum of EC~002 \citep{galinier2023}. Such a band is not as clearly observable in the VIS spectrum of (14390) 1990~QP10, but this object shows a flatter slope than V-types in the 500 - 750 nm region \citep{Matlovic2020}, which could be indicative of a weak absorption.

After analysis of their spectral features and comparison with taxonomic end-members, these three asteroids have been found not belonging to any known literature spectral types \citep{moskovitz2008_10537,Duffard2009,leith2017,Hardersen2018,Migliorini2018}. \cite{Migliorini2018} even suggest that asteroid (14390) 1990~QP10 might be a unique example of a new taxonomic class. Given their absorption features, \cite{Hardersen2018} and \cite{moskovitz2008_10537} suggest that these objects could be composed of a mixture of pyroxenes and olivine; and \cite{moskovitz2008_10537} suppose that asteroid (10537) 1991~RY16 could be an isolated fragment of a partially or fully differentiated parent body. Given their peculiar spectral characteristics and their lack of a satisfactory match with known asteroids or meteorites, these objects constitute candidate analogues of EC~002.

In the following, we present in Sect.\ref{data} the observations in the NIR we performed to validate or not the potential analogues of EC~002 found in \cite{galinier2023}, and other data used for the present study. The spectral analysis of the asteroids and the results are presented in Sect.\ref{analysis}, and a discussion is provided in Sect.\ref{sec:discussion}.


\section{Data}
\label{data}

\subsection{Observations}

Between August 2023 and January 2024, we acquired 21 NIR spectra of 20 different asteroids using NASA's Infrared Telescope Facility (IRTF), with asteroid (24286) 1999 XU188 which was observed twice. Of the observed asteroids, 12 were identified by \cite{galinier2023} as matching the powder spectrum of EC~002; seven matched the spectrum modelled with low space weathering, and one matched the spectrum modelled with medium space weathering of the meteorite. The complete list of observed objects is presented in Appendix \ref{app:obs} Table~\ref{tab:obs}, along with the observation date and some associated parameters. The observed spectra are publicly available on the IRTF Data Archive \footnote{IRTF Data Archive: \url{https://irtfweb.ifa.hawaii.edu/research/irtf_data_archive.php}}.



Among the instruments mounted on the IRTF, SpeX and MORIS were used for our observations. SpeX is a NIR spectrograph and imager that we used in the low resolution prism mode (R $\simeq$ 200) to characterise our faint targets \citep{marsset2020,rayner2003}. This mode spans the wavelength range from 700 to 2520 nm, allowing the spectroscopic characterisation of the surface composition of asteroids. We used a spectroscopic slit of 0.8 x 15 arcsec. We used as well a dichroic reflector of 0.7 nm to redirect the light of faint targets from the spectrograph to the CCDs of the guiding camera MORIS \citep{Gulbis2011}, which allows to guide the telescope on objects having V magnitudes as low as 20. The spectra of asteroids and of solar analogues stars were acquired in ABBA pairs of exposures, and calibration frames were acquired once a night for bias and flat-field corrections \citep{Popescu2014}.

The observations were reduced using Spextool, a spectral reduction tool provided by the IRTF which allows the production of reflectance spectra after flat-field correction and wavelength calibration \citep{rayner2003}. The data reduction was done as in \cite{Avdellidou2022}: each observed raw asteroid spectrum was divided by a corresponding solar analogue spectrum, to obtain a reflectance spectrum, which was then normalised. The spectrum was then corrected for telluric lines. The asteroid reflectance spectrum $R(\lambda)$ was thus obtained using the following equation:
\begin{equation}
	R(\lambda) = \frac{A(\lambda)}{S_L(\lambda)} \times 
	Poly \left (\frac{S_L(\lambda)}{S_T(\lambda)} \right),
	\label{E:spec}
\end{equation}
where $A(\lambda)$ is the wavelength-calibrated raw spectrum of the asteroid, $S_L(\lambda)$ is the wavelength-calibrated raw spectrum of the local G2V star observed within $\sim$300" of the asteroid, and $S_T(\lambda)$ is the wavelength-calibrated raw spectrum of the selected well-studied solar analogue star observed at a similar airmass to the asteroid. In general, the local solar analogue star guarantees the accuracy of telluric line removal, but the difference between the spectrum of the local star and that of the Sun may require an additional correction of the spectrum's slope. This correction is defined as $Poly \left (S_L(\lambda)/S_T(\lambda) \right)$, the $Poly()$ function representing a polynomial fit of the ratio of the stars spectra. The regions of the spectra affected by telluric features, corresponding to 1300 $< \lambda <$ 1500, 1780 $ < \lambda <$ 2100, and $\lambda > $ 2400 nm, were excluded.  Finally, each asteroid's spectrum was shifted to sub-pixel accuracy, for it to be aligned with the calibration star spectra. In case the local G2V star was not observed, the asteroid's spectrum was directly divided by a well-studied solar analogue's spectrum, using $R(\lambda)=  A(\lambda)/S_T(\lambda)$, which insures the correction of telluric lines \citep{Avdellidou2022}.

\subsection{Literature data}

Along with our observations and the laboratory and space-weathered modelled spectra of EC~002 of \cite{barrat2021}, we used the dataset of VIS reflectance spectra of 60,518 Solar System objects (SSOs) released in the Gaia DR3 \citep{galluccio2022} to perform our analysis. These spectra cover the wavelength range from 374 to 1034 nm in 16 bands, and were produced by merging spectra acquired independently by the Blue and Red spectrophotometers (BP and RP) onboard Gaia, spanning the respective wavelength ranges [325, 650] nm and [650, 1125] nm \citep{galluccio2022}. The Gaia DR3 reflectance spectra are normalised at 550 nm, and to each wavelength band is associated a `spectral\_validation\_flag' (hereafter flag) number, assessing the estimated quality of the band: flag=0 for good quality, 1 for lower quality or suspected issue, and 2 for bad quality \citep[see][for details]{galluccio2022}.

The near-infrared spectrum of (31060) 1996~TB6, that was identified as a potential spectral match to EC~002 in \cite{galinier2023}, has already been published and taxonomically classified in \cite{Delbo2026}. In this study, we use the same spectrum to carry out our analysis, and we treat this object as part of the observational campaign in the following. The VISNIR spectrum of asteroid (10537) 1991~RY16 of \cite{moskovitz2008_10537} was used as well, along with the NIR spectra of asteroids (7472) Kumakiri and (14390) 1990 QP10 \citep{leith2017}.




\section{Analysis and results}
\label{analysis}

\subsection{Classification of observed asteroids}
\label{sec:class}

To analyse the observed asteroids and compare them with EC~002, we combined their VIS Gaia DR3 spectra with the observed NIR spectra to produce full VISNIR spectra. To do so, Gaia DR3 spectra were considered only between 462 and 946 nm, excluding the first and last two spectral bands as they are often affected by systematic issues \citep{galluccio2022,galinier2023}. The observed NIR spectra were resampled between 690 and 2400 nm using 5 nm steps, and were normalised at 1000 nm thanks to the calculation of a normalisation factor, which was then applied to the original NIR spectra. Next, the normalised NIR and Gaia DR3 spectra of each observed asteroid were merged by applying a multiplicative factor $\alpha$ to the NIR spectrum, to scale it with its corresponding Gaia DR3 spectrum normalised at 550 nm. The factor $\alpha$ was chosen depending on the satisfaction of the visual scaling of the VIS and NIR spectra, taking into account a potential reddening of the last bands of the Gaia DR3 spectra \citep{galluccio2022,galinier2023,Oszkiewicz2023}. The scaled spectra are displayed in Appendix \ref{app:obs} Fig.~\ref{fig:EC002_analogues_nir}. In some cases the factor $\alpha$ was left to unity, when the scaling was visually satisfying -- for asteroid (1643) Brown for example -- or when no value of the factor appeared to improve the scaling -- for asteroid (12551) 1998 QQ39 for example. The compared results of this visual scaling and of an automatic scaling method is discussed in Appendix \ref{app:visir_merging}.


We can see in Appendix \ref{app:obs} Fig.~\ref{fig:EC002_analogues_nir} that the spectra of (12551) 1998 QQ39 and (27884) 1996 EZ1 show an unexpected discontinuity between their VIS and NIR parts: their NIR spectrum is flat and red in the overlapping region with the VIS spectrum, which is a V-type kind of spectrum with a blue slope longward 700 nm. Since there is no clear explanation for this spectral discontinuity, we discarded these two asteroids from the analysis.

We classified the remaining 19 VISNIR spectra of the observed asteroids in the Bus-DeMeo taxonomic scheme, using the online MIT classification tool\footnote{http://smass.mit.edu/busdemeoclass.html} of \cite{demeo2009}. To do so, we smoothed and sampled the spectra in 41 data points between 450 and 2450 nm using a 50 nm interval, as required by the classification algorithm; and we normalised the resampled spectra at 550 nm.

The best class of each asteroid given by the classification algorithm is given in Appendix \ref{app:obs} Table.\ref{tab:obs}. Most asteroids are classified as V-types, a spectral class characteristic of a basaltic composition. Asteroids classified as Vw are V-types with slopes greater than 0.25, as defined in \cite{demeo2009}. These objects could be red-sloped V-types, or could be assigned the w suffix here because of a potential reddening of the last bands of Gaia DR3 spectra \citep{galluccio2022,galinier2023,Oszkiewicz2023}. Such reddening could lead to an increase of the calculated slope of the VISNIR spectra after merging the VIS with the NIR part. However, such potential reddening does not impact the classification of these objects as V-types. Asteroid (18780) Kuncham was found to belong to the S-complex, while asteroid (31060) 1996 TB6 is classified A-type, therefore olivine-rich, as found by \cite{Delbo2026}.  



\subsection{Spectral features analysis: Band I Centre and Band Area Ratio}

In the following, we analyse the diagnostic spectral features representative of the composition of asteroids: the presence and position of the band around 1 $\mu$m (Band I centre BIC), and the Band Area Ratio (BAR). The band centres are directly related to the mineral composition of an object \citep{Gaffey2011}, and these features were found to be insensitive to space weathering for spectra presenting strong features \citep{Marchi2005,Gaffey2010,brunetto2015}. We did not correct for the influence of temperature on these spectral features \citep{Moskovitz2010,Matlovic2020}, since the effect of temperature on the BIC is negligible \citep{Burbine2009,Sanchez2012} and we do not know at which temperature the meteorite spectra were acquired. We did not consider the position of the band around 2 $\mu$m (Band II centre BIIC) or the band depths here, as the BIIC was found to be more affected by temperature than the BIC \citep{Burbine2009,Sanchez2012}, and band depths can be significantly affected by space weathering \citep{Fu2012}. Moreover, some of our observations are strongly affected by telluric absorption, which may make it difficult to determine the band centre and depth. Besides, the BIC vs BAR space is particularly significant and relevant for meteorite - asteroid comparison \citep{Moskovitz2010}.


\begin{figure}[!h]
	\centering 
	\begin{adjustbox}{clip,trim=0.cm 0cm 0.cm 0.4cm,max width=\textwidth} 
		\includegraphics[scale=0.6]{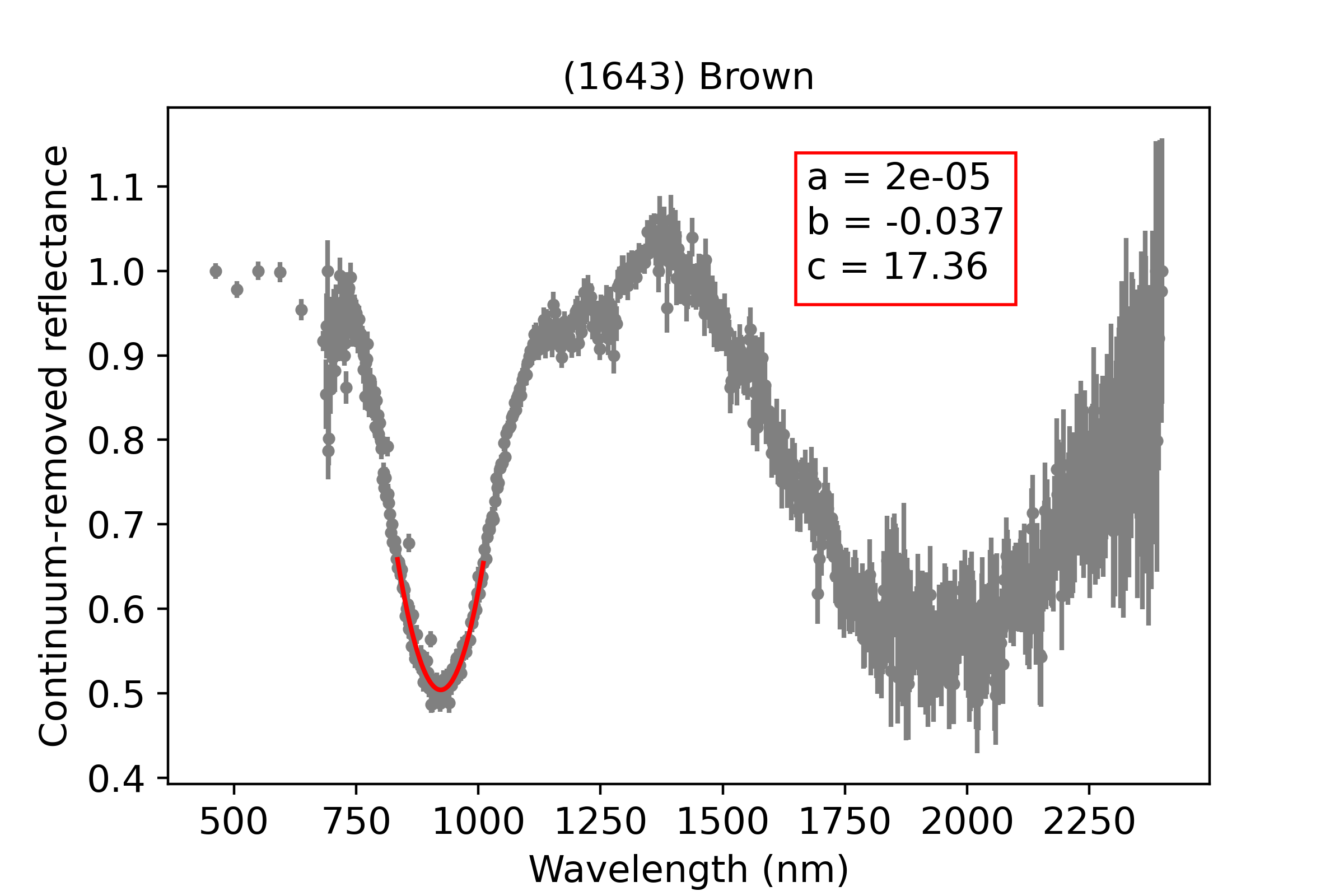} 
	\end{adjustbox}
	\caption{Second order polynomial fit of the 1 $\mu$m band on the continuum-removed spectrum of (1643) Brown. The polynomial fit parameters are displayed inside the red square, the equation of the fit being $y=ax^2+bx+c$.}
	\label{fig:band_fit}
\end{figure}

We calculated the BIC and BAR for the 19 VISNIR asteroid spectra that do not show a discrepancy between their Gaia DR3 and observed NIR spectra. The wavelengths longwards 2400 nm were not taken into account, as they are affected by telluric absorptions \citep{marsset2020,Avdellidou2022}. The BIC is located at the maximum absorption of the 1 $\mu$m feature. To calculate it, we removed the continuum of each spectrum performing a convex hull fit (\textit{scipy} ConvexHull function) to the full VISNIR spectra, assuming a linear continuum \citep{Gaffey2011,Fu2012}. We did not consider regions affected by telluric absorption around 1400 and 1800 nm in the calculation of the continuum, which would result in an error in the calculation of the continuum for heavily affected spectra, such as that of (3869) Norton (see Fig.~\ref{fig:EC002_analogues_nir}). Then, the convex hull was interpolated at the original wavelengths of the spectra, and each spectrum was divided by their calculated continuum to produce continuum-less reflectance spectra. 


We followed a method similar to the one described in \cite{Storm2007} to calculate the BIC from the continuum-less reflectance spectra. Considering that the first band spans the wavelength range from 700 to 1200 nm, we performed a second degree polynomial fit over the bottom third of this band \citep{Storm2007}, using nympy's \textit{polyfit} function. The BIC was then calculated using the parameters of the polynomial fit: considering that it is described by the equation $y=ax^2+bx+c$, the BIC was determined as the band minimum of the continuum-less spectra:
\begin{equation}
	BIC=\frac{-b}{2a}
\end{equation}
An example of band fitting is shown in Fig.\ref{fig:band_fit} for (1643) Brown. The parameters of the fit are displayed in the red square in the figure.

Using a Monte Carlo method, we then randomly resampled each reflectance value in the bottom third of the 1 $\mu$m band 999 times, using a normal distribution, to describe the uncertainty associated with each reflectance point \citep{Storm2007}. We calculated the BIC associated to these reflectance values using the previous formula, and then we deduced the mean value and standard deviation of the 1000 measurements of the band centre. The average BIC value and associated uncertainty obtained for each asteroid spectrum are presented in Table.\ref{tab:BIC_BAR_ast}.

\begin{figure}[!h]
	\centering 
	\begin{adjustbox}{clip,trim=0.5cm 0.1cm 0.5cm 0.4cm,max width=\textwidth} 
		\includegraphics[scale=0.4]{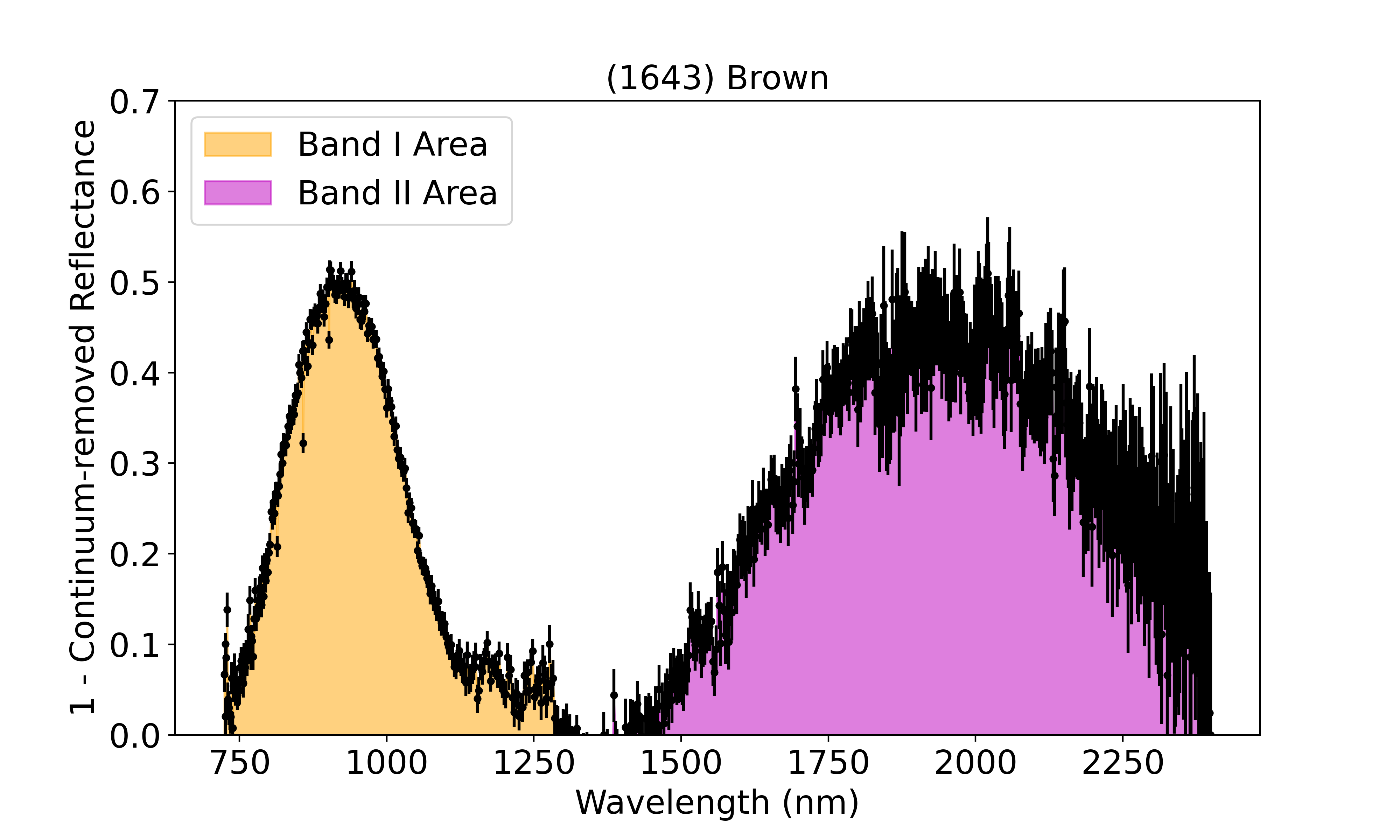} 
	\end{adjustbox}
	\caption{Band areas of the first and second bands of the continuum-removed spectrum of (1643) Brown.}
	\label{fig:BAR}
\end{figure}

The BAR is defined as the area of Band II divided by the area of Band I. To determine it, we calculated the opposite of the continuum-less reflectance spectra, to then calculate the area under each band using the trapezoidal method, integrating along the parametric curve (using \textit{numpy}'s \textit{trapz} function). The wavelength range spanned by each band changes depending on the asteroid spectral type. For V-types, we defined the limit between the two bands $\lambda_{lim}$ to be around 1350 nm. For the A-type asteroid (31060) 1996 TB6, the limit between the two bands was defined at 1700 nm; and for S-complex asteroid (18780) Kuncham, the limit was defined at 1500 nm. A figure illustrating the calculation of the two band areas for asteroid (1643) Brown is displayed in Fig.\ref{fig:BAR}.

Same as for the BIC and following the method described in \cite{Dunn2013}, we used a Monte Carlo method to vary the reflectance values within their uncertainties 999 times, considering a normal distribution, and we calculated the average BAR and its associated uncertainty for each asteroid. The calculated BAR averages and their uncertainty for each asteroid spectrum considered are shown in Table.\ref{tab:BIC_BAR_ast}. 

\begin{table}[!h]
	\centering
	\caption{Average BIC and BAR and associated uncertainty calculated for every considered asteroid, and BIC and BAR values associated with the laboratory spectra of EC~002. (a) and (b) refer to the two observed spectra of the asteroid (24286) 1999 XU188. \label{tab:BIC_BAR_ast}}
	\begin{tabular}{l c c}
		\hline
		Observed asteroid & BIC (nm) & BAR \\ 
		\hline 
		(1643) Brown  &  923.1 $\pm$ 0.8 & 2.16 $\pm$ 0.03 \\
		(1946) Walraven  &  927.9 $\pm$ 0.7 & 1.92 $\pm$ 0.02 \\
		(3188) Jekabsons  &  923.1 $\pm$ 0.3 & 2.02 $\pm$ 0.01 \\
		(3869) Norton  &  930.7 $\pm$ 1.6 & 2.90 $\pm$ 0.07 \\
		(4302) Markeev  &  926.0 $\pm$ 0.4 & 2.20 $\pm$ 0.01 \\
		(6003) 1988 VO1  &  923.2 $\pm$ 1.3 & 3.69 $\pm$ 0.07 \\
		(6789) Milkey  &  925.1 $\pm$ 0.6 & 2.18 $\pm$ 0.02 \\
		(8827) Kollwitz  &  933.9 $\pm$ 0.7 & 2.02 $\pm$ 0.02 \\
		(10671) Mazurova  &  937.9 $\pm$ 1.8 & 2.32 $\pm$ 0.05 \\
		(14511) Nickel  &  923.0 $\pm$ 0.6 & 2.60 $\pm$ 0.03 \\
		(15989) Anusha  &  933.6 $\pm$ 0.6 & 2.06 $\pm$ 0.02 \\
		(17056) Boschetti  &  938.1 $\pm$ 0.6 & 1.92 $\pm$ 0.02 \\
		(18780) Kuncham  &  1019 $\pm$ 11.7 &  1.02 $\pm$ 0.03 \\
		(20454) Pedrajo  &  927.8 $\pm$ 0.8 & 2.00 $\pm$ 0.02 \\
		(24286) 1999 XU188 (a)  & 930.3 $\pm$ 0.5 & 2.46 $\pm$ 0.02  \\
		(24286) 1999 XU188 (b)  & 927.5 $\pm$ 0.8 & 1.98 $\pm$ 0.02  \\
		(30426) Philtalbot  &  934.6 $\pm$ 1.3 & 2.56 $\pm$ 0.04 \\
		(31060) 1996 TB6  & 1067.0 $\pm$ 1.1 &  0.23 $\pm$ 0.01 \\
		(45787) 2000 OJ24  &  921.3 $\pm$ 1.0 & 2.60 $\pm$ 0.03 \\
		\hline
		Literature asteroid & BIC (nm) & BAR \\ 
		\hline
		(10537) 1991~RY16 & 968.8 $\pm$  1.4 & 0.98 $\pm$  0.02  \\ 
		(7472) Kumakiri & 1022.9 $\pm$  1.0 & 0.30 $\pm$ 0.01 \\ 
		(14390) 1990~QP10 &  999.8 $\pm$  3.0 & 0.90 $\pm$  0.03 \\ 
		\hline
		EC~002 sample & BIC (nm) & BAR \\ 
		\hline
		Powder  &  983.3 & 0.96 \\
		slab 1  &  987.3 & 1.15 \\
		slab 2  &  986.9 & 1.05 \\
		slab 3 &  985.9 & 1.16 \\
		\hline
	\end{tabular}
\end{table}

\begin{figure}[!htb]
	\centering
	\begin{subfigure}{0.49\textwidth}
		\centering
		\begin{adjustbox}{clip,trim=0.5cm 0.2cm 0.5cm 0.5cm,max width=\linewidth}
			\includegraphics{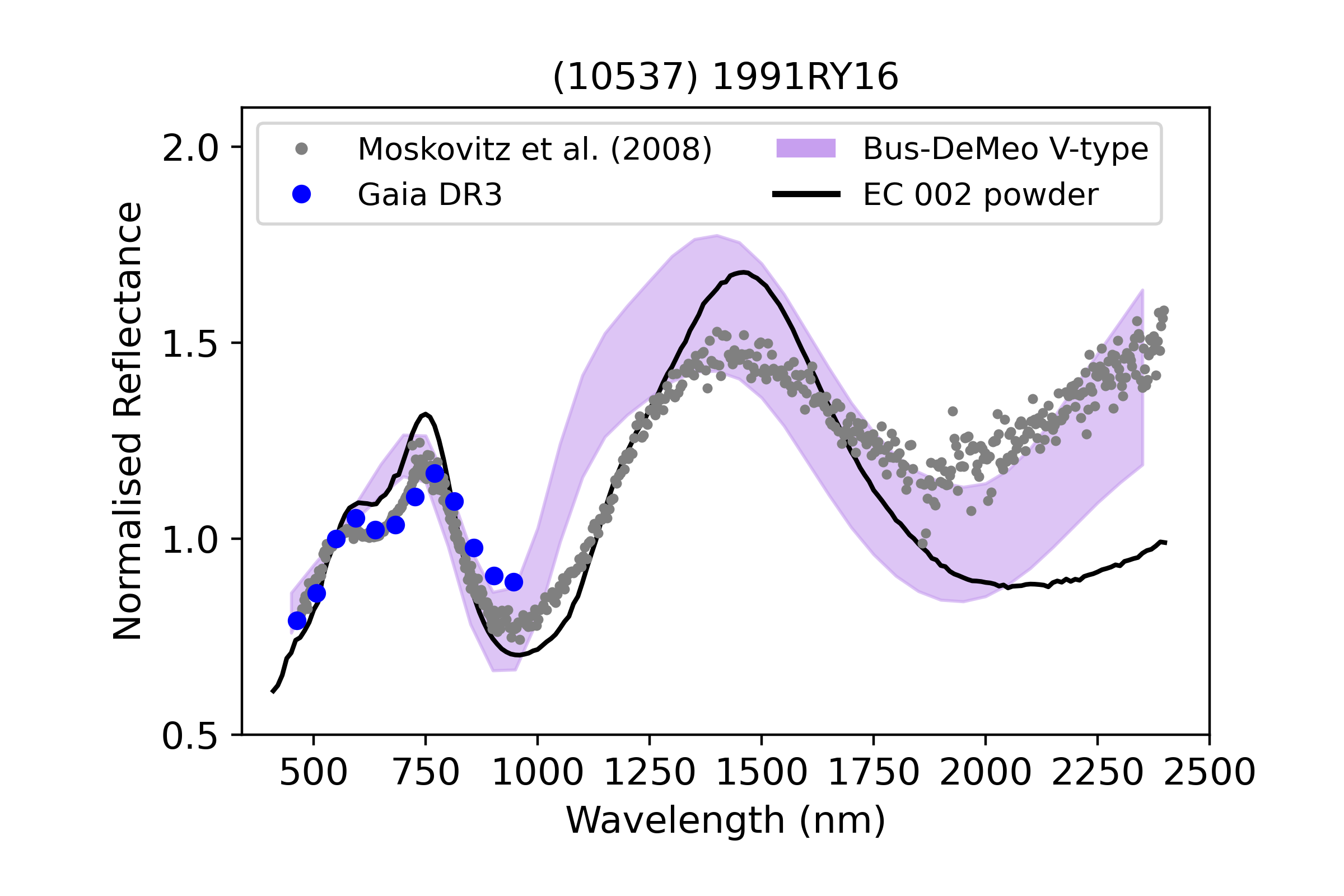}
		\end{adjustbox}
	\end{subfigure}
	\hfill
	\begin{subfigure}{0.49\textwidth}
		\centering
		\begin{adjustbox}{clip,trim=0.5cm 0.2cm 0.5cm 0.5cm,max width=\linewidth}
			\includegraphics{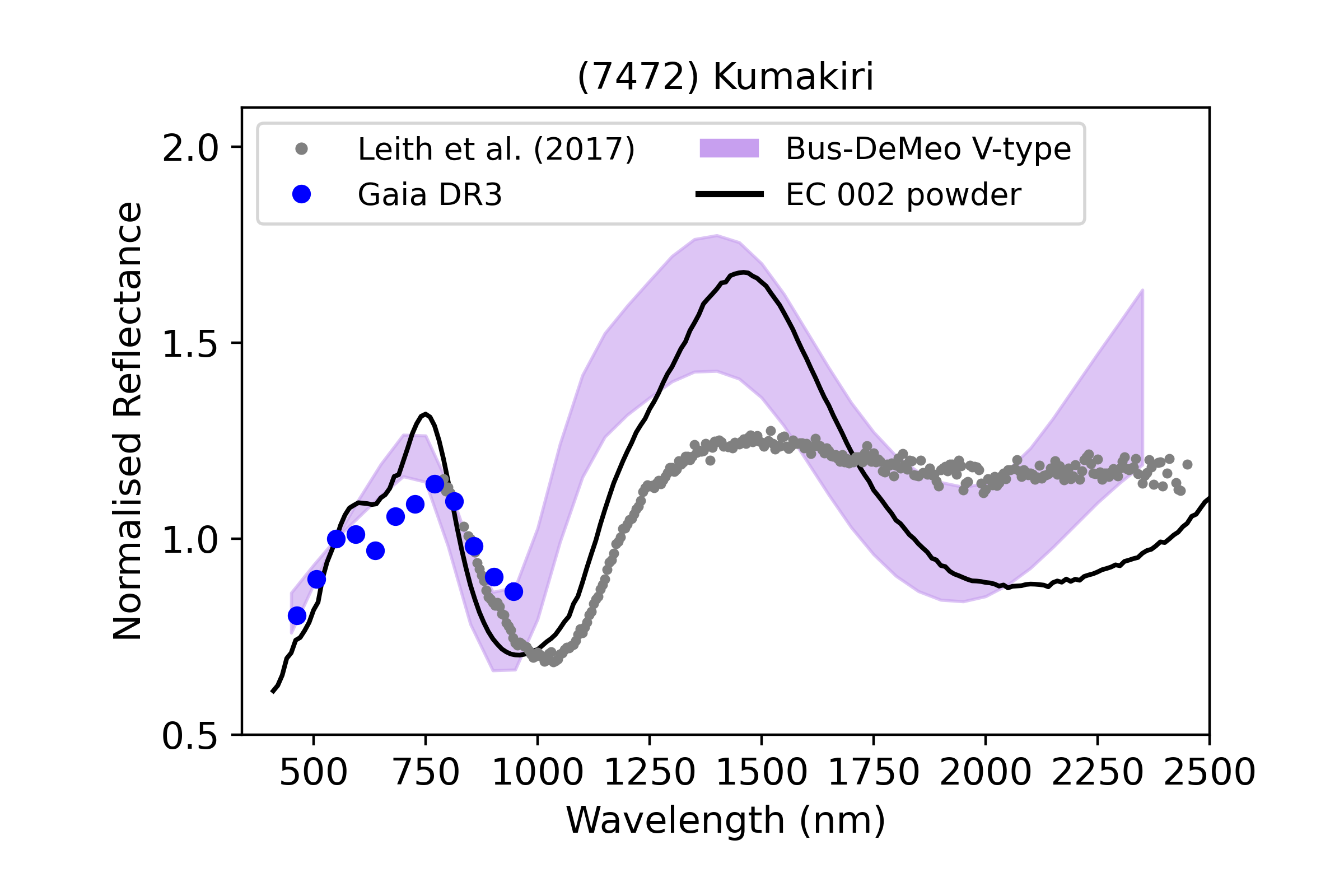}
		\end{adjustbox}
	\end{subfigure}
	\hfill
	\begin{subfigure}{0.49\textwidth}
		\centering
		\begin{adjustbox}{clip,trim=0.5cm 0.5cm 0.5cm 0.5cm,max width=\linewidth}
			\includegraphics{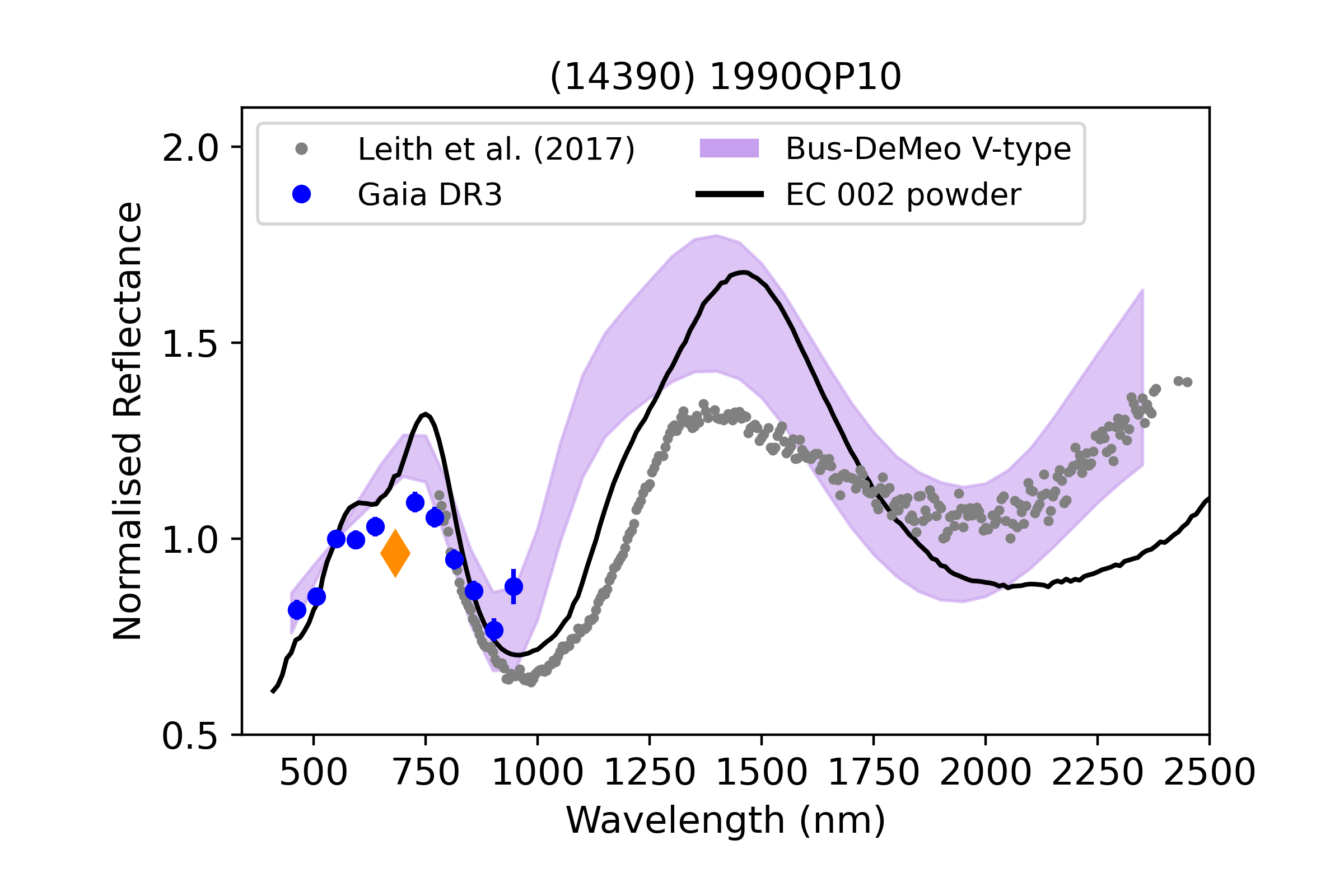}
		\end{adjustbox}
	\end{subfigure}
	\caption{Literature spectra of asteroids (10537) 1991~RY16  \citep{moskovitz2008_10537}, (7472) Kumakiri  \citep{leith2017} and (14390) 1990~QP10  \citep{leith2017} (grey dots), and Gaia DR3 spectra considering the range from 462 to 946 nm (blue dots for bands with flag=0, orange diamond for bands with flag=1). The V-type template spectrum of the Bus-DeMeo taxonomic scheme \citep{demeo2009} is shown as a purple shaded area, and the powder sample's spectrum of meteorite EC~002 is displayed in black lines for comparison.}
	\label{fig:lit_weird}
\end{figure}

Then, the BIC and BAR parameters were calculated on the full VISNIR spectrum of asteroid (10537) 1991~RY16 of \cite{moskovitz2008_10537}, and on the spectra of asteroids (7472) Kumakiri and (14390) 1990~QP10 \citep{leith2017}. The VISNIR spectra of the latter objects were produced by bridging their VIS Gaia DR3 spectrum normalised at 550 nm with their literature NIR spectrum, using an $\alpha$ factor of 0.7. The VISNIR spectra of these three asteroids are displayed in Fig.\ref{fig:lit_weird}, along with the V-type template spectrum of the Bus-DeMeo taxonomy \citep{demeo2009} and the powder sample spectrum of EC~002 for comparison.


We did not use the values for the BIC and BAR parameters reported in \cite{moskovitz2008_10537}, in \cite{leith2017} or in \cite{Hardersen2018} for these objects, as they were calculated using different techniques. We preferred using an internally consistent method to compare our spectra, to avoid introducing differences due to these slightly different techniques \citep{Moskovitz2010,Duffard2006}. We calculated the BIC and BAR values using the same method as the one applied to the observed asteroids, setting the wavelength limit to calculate the BAR to 1400 nm. The obtained values for the spectral features are given in Table.\ref{tab:BIC_BAR_ast}.

Finally, in order to compare the meteorite with observed asteroids, the BIC and BAR parameters were calculated for the powder and three slabs spectra of EC~002. We used the same method as for asteroids, but we neglected the error bars associated with the meteorite's spectra, as they are much smaller than those associated with the asteroids' spectra \citep{Moskovitz2010}. We therefore did not need to use a Monte Carlo resampling to calculate uncertainties for these spectra. The wavelength corresponding to the limit between Band I and Band II and needed to calculate the BAR $\lambda_{lim}$ was defined at 1450 nm. The BIC and BAR values obtained for each sample are given in Table.\ref{tab:BIC_BAR_ast}.

\begin{figure}[!h]
	\centering 
	\includegraphics[scale=0.35]{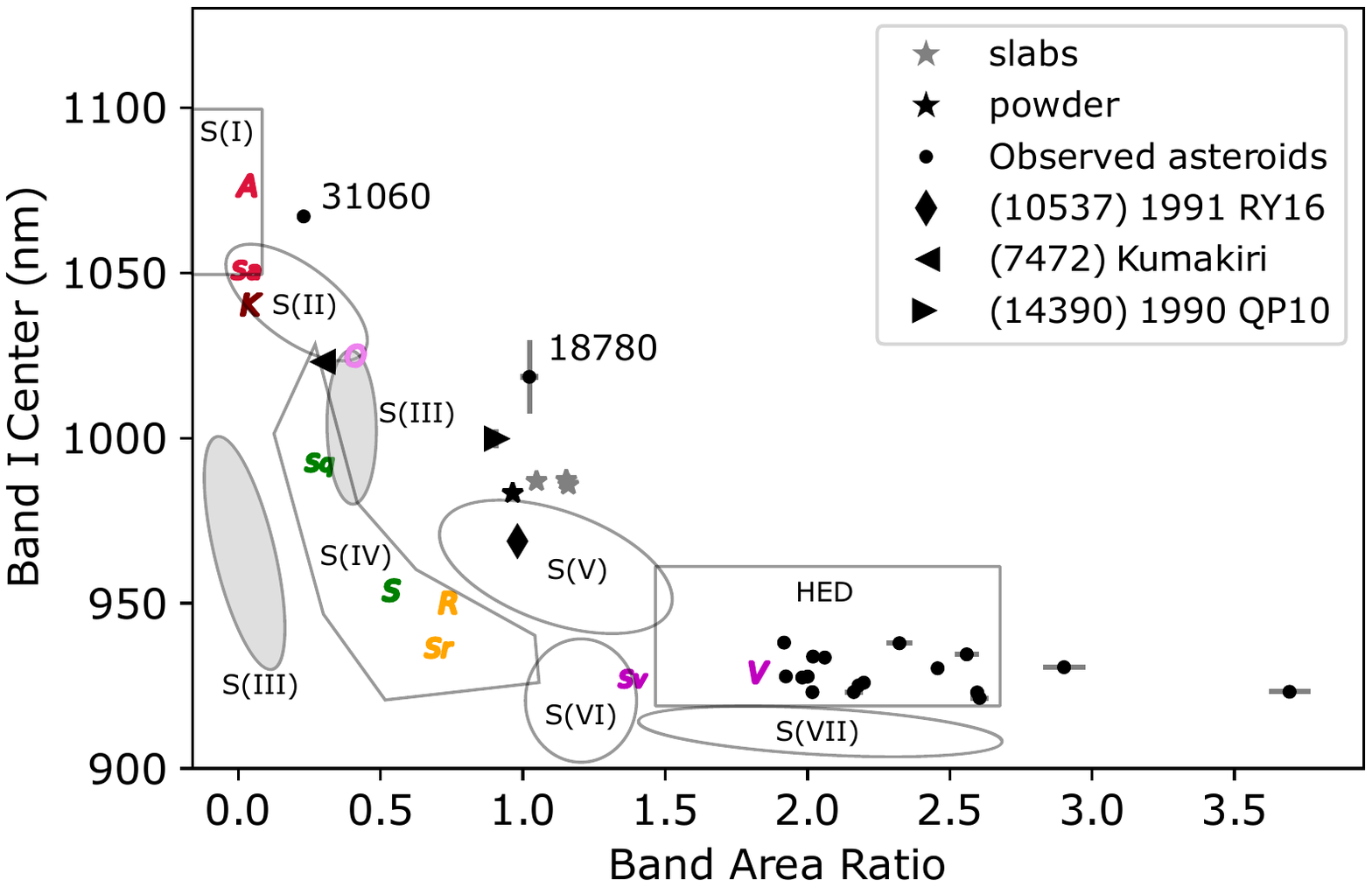} 
	\caption{Band I Centre (BIC) as a function of the BII/BI band area ratio (BAR). The BIC and BAR of the meteorite powder and slab spectra are displayed as black and grey stars respectively, and the observed asteroids as black dots with their associated uncertainty. The identification number of S-complex asteroid (18780) Kuncham and A-type asteroid (31060) 1996~TB6 are written next to their corresponding data point. The BIC and BAR of the taxonomic end-members of \cite{demeo2009} are highlighted in colours, the letter corresponding to the taxonomic type of the end-member. Finally, asteroid (10537) 1991~RY16 is plotted as a black diamond and (7472) Kumakiri and (14390) 1990~QP10 as black triangles. The mineralogical subdivisions defined by \cite{Gaffey2011} for S-type asteroids, ordinary chondrites, basaltic achondrites, and olivine assemblages are overlaid in light grey.}
	\label{fig:BIC_BAR}
\end{figure}

The respective position in the BIC vs BAR plot of the observed asteroids, of EC~002 laboratory spectra, and of asteroids (10537) 1991~RY16, (7472) Kumakiri and (14390) 1990~QP10 are displayed in Fig.\ref{fig:BIC_BAR}. We highlighted in this figure the BIC and BAR calculated for some of the taxonomic end members of the Bus-DeMeo classification presenting spectra with strong features, following what was shown in the supplementary material of \cite{barrat2021} (see their Fig. S13). On this plot, we can observe that the parameters derived for EC~002 plot quite far from those of the taxonomic end-members, as noted by \cite{barrat2021}. Most of the asteroids we observed show parameters matching with basaltic V-type objects, which is coherent with their taxonomic type determined in Sect.\ref{sec:class}. The majority of these V-types are inside the mineralogical area of HED meteorites defined by \cite{Gaffey2011}, indicative of their basaltic composition, as expected from their spectral type. Two V-types plot outside of this area: (3869) Norton and (6003) 1988 VO1. These objects are the most heavily affected by the telluric absorption around 1300 nm, which can result in an overestimation of their BII area and thus of their BAR, explaining their position in the BIC vs BAR space. In any case, we can deduce from their position in this space that the 16 V-type asteroids we observed have a composition closer to that of HED meteorites than to that of EC~002. Asteroid (31060) 1996~TB6 plots in the A-type region, as expected from its spectral class \citep{Delbo2026}, which is consistent with a high content in olivine \citep{Gaffey2010}. All these objects are located far from the meteorite in this space. 


\begin{figure}[h!]
	\centering 
	\begin{adjustbox}{clip,trim=0.cm 0.cm 0.cm 0.4cm,max width=\textwidth} 
		\includegraphics[scale=0.6]{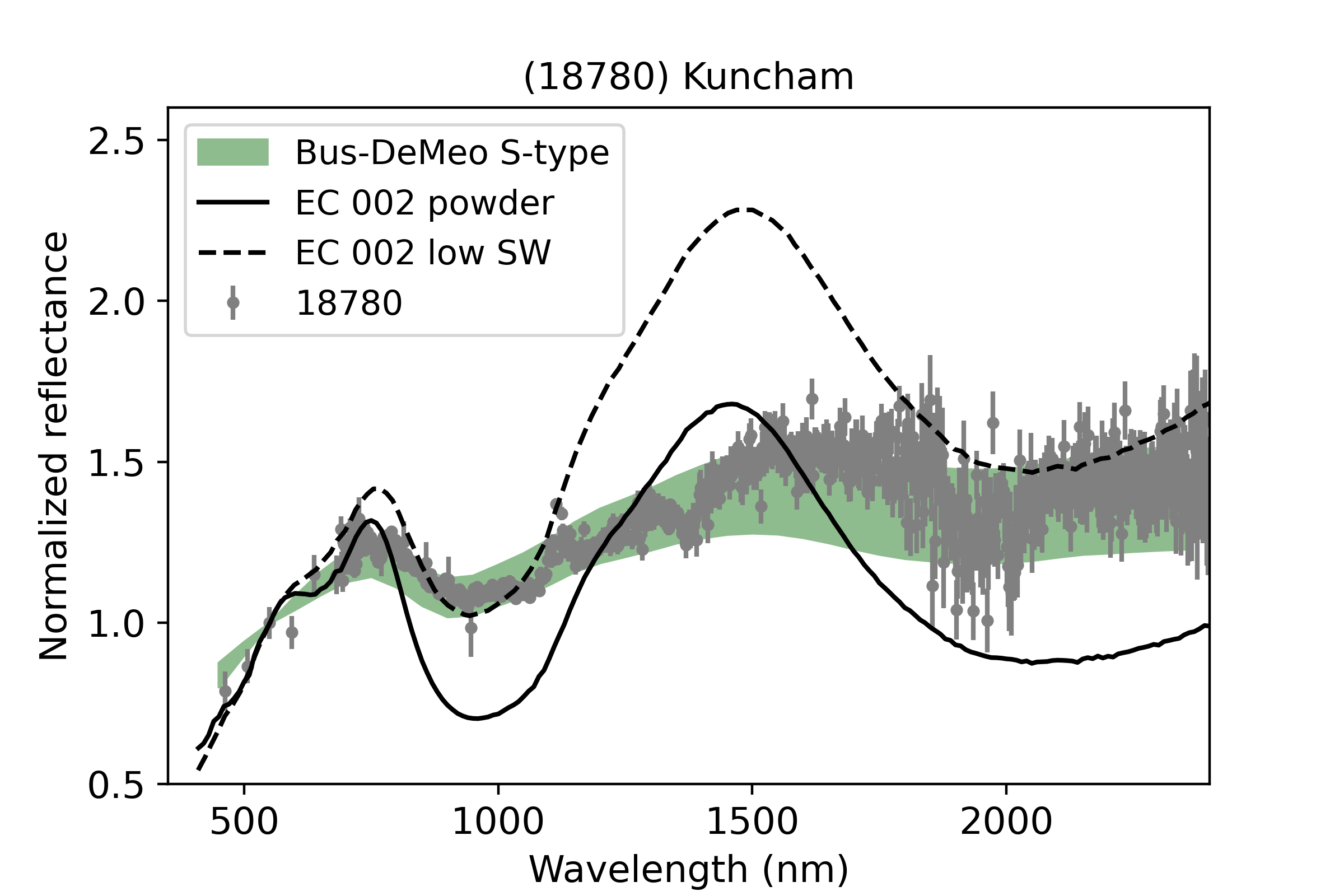} 
	\end{adjustbox}
	\caption{Spectrum of asteroid (18780) Kuncham (grey dots) superimposed with the S-type template spectrum of the Bus-DeMeo taxonomy \citep{demeo2009} (green shaded area). The powder sample and low space-weathered modelled spectra of meteorite EC~002 are displayed for comparison.}
	\label{fig:18780_S}
\end{figure}

The asteroid we observed that is closest to the meteorite in this space is (18780) Kuncham. However, this object was observed because it potentially matched the spectrum of EC~002 modelled with low space-weathering, and it has noisy Gaia DR3 and NIR spectra. In addition, it is quite strongly affected by telluric absorptions in the NIR, as can be seen in Fig.\ref{fig:18780_S}, which can result in an overestimation of its BAR . Its spectrum is not compatible with the laboratory and low space-weathered spectra of EC~002, and it is therefore unlikely to be a good match for the meteorite. 

Finally, in the BIC vs BAR space, asteroids (10537) 1991~RY16 and (14390) 1990~QP10 plot the closest to the meteorite spectra, which makes them good matches for EC~002. Their position, far from any taxonomic end-members, is consistent with Fig.2 of \cite{moskovitz2008_10537} and further confirms the unique character of these asteroids spectra \citep{moskovitz2008_10537,Duffard2009,leith2017,Hardersen2018,Migliorini2018}. Asteroid (7472) Kumakiri plots closer to the taxonomic end-member O-type, as noted by \cite{Hicks2014}. 

\section{Discussion}
\label{sec:discussion}

We exploited the observation of 20 asteroids identified as potential analogues of EC~002 in \cite{galinier2023}, for a total of 21 spectra. We classified 16 of these objects as being of the basaltic V-type, one as being of the olivine-rich A-type as in \cite{Delbo2026}, and another as belonging to the S-complex. Two objects we observed were not included in the analysis because they show an unexplained discrepancy between their Gaia VIS spectrum and their acquired NIR spectrum.


\begin{figure*}[!h]
	\centering
	\begin{adjustbox}{clip,trim=0.cm 0.5cm 0.5cm 0.5cm,max width=\textwidth}
		\input{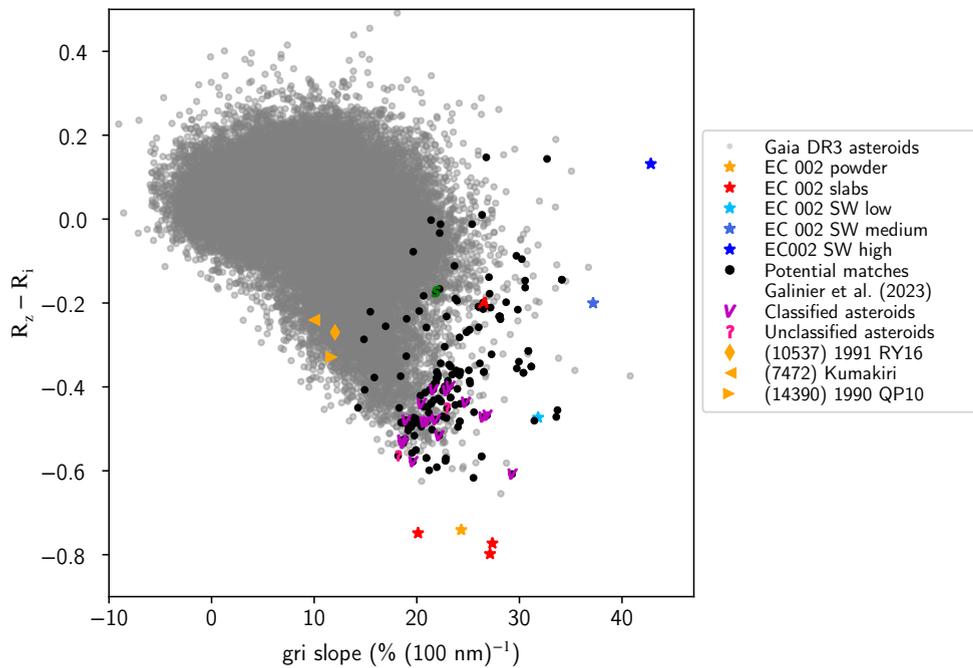}
	\end{adjustbox}
	\caption{Distribution of the depth of the band around 1000 nm with respect to the gri spectral slope of every Gaia asteroid (grey dots), see \cite{galinier2023}. Red stars: raw slabs; orange star: powder sample of EC~002. Stars going from light blue to dark blue: modelled space-weathered spectra of EC 002, with different space weathering intensity. Black dots: potential analogues of EC~002 determined in \cite{galinier2023}. The asteroids observed and classified in this paper are represented by a letter indicating their spectral type: magenta V-types, red A-type and green S-type. The two asteroids (12551) 1998 QQ39 and (27884) 1996 EZ1 are shown as pink interrogation marks, as their NIR spectra did not allow a reliable classification. Asteroid (10537) 1991~RY16 is highlighted as an orange diamond and (7472) Kumakiri and (14390) 1990~QP10 as orange triangles.}
	\label{fig:slope_zi}
\end{figure*}

By comparing the BIC and BAR parameters of the observed objects with those of the meteorite samples, none of the 20 asteroids we observed appears to match the meteorite's spectrum. However, we observed only 20 objects out of the 142 potential matches of EC~002 determined in \cite{galinier2023}. To determine the probability to observe a matching asteroid among these 142 objects knowing that none of the 20 asteroids we observed is a match, we used a Bayesian beta-binomial model assuming a uniform prior distribution. We determined that the probability that the next asteroid we could observe among the 142 potential matches is a successful match to the meteorite is only of 4.5$\pm$4\%, which is low. This result can have several explanations. 

Firstly, it is difficult to distinguish the spectrum of EC~002 from that of a basaltic body in the VIS wavelength range, which explains why the selection of potential analogues made in \citep{galinier2023} is mainly composed of V-type objects. Moreover, the `fake band' issue affecting some Gaia DR3 spectra can further cause a basaltic V-type asteroid to appear as a potential analogue of the meteorite, as mentioned in the introduction. Indeed, out of the 20 observed asteroids, 16 have been predicted by \cite{Oszkiewicz2023} to be V-types from the study of their Gaia DR3 spectra, using various machine learning methods. We confirm here that asteroids (1643) Brown, (3188) Jekabsons, (4302) Markeev, (6003) 1988 VO1, (6789) Milkey, (8827) Kollwitz, (10671) Mazurova, (14511) Nickel, (15989) Anusha, (17056) Boschetti, (20454) Pedrajo, (24286) 1999 XU188, (30426) Philtalbot, and (45787) 2000 OJ24 are V-types. Asteroid (1946) Walraven shows three bands having a non-zero flag in the Gaia DR3, and was therefore not included in the sample of objects studied by \cite{Oszkiewicz2023} since the authors required spectra with flag=0 for all used wavelengths in their selection. Asteroid (3869) Norton was not characterised either, and we cannot conclude on (12551) 1998 ~QQ39 and (27884) 1996~EZ1 from their NIR spectra. Among the observed objects, four are found to be (4) Vesta family members in several catalogues \citep{nesvorny2015,milani2014,Vinogradova2019,Broz2013}: asteroids (6003) 1988 VO1, (20454) Pedrajo, (27884) 1996~EZ1 and (45787) 2000 ~OJ24. Their classification as V-types, except for (27884) 1996~EZ1, is consistent with their belonging to the Vesta family. V-types are further discussed in Appendix \ref{app:V}.

We can note that the spectrum of (27884) 1996 EZ1 is similar to the one of (11699) 1998 FL105 of \cite{Hardersen2018}, both objects displaying spectra with a steep red slope in the near-infrared. However, the spectrum of (11699) 1998 FL105 of \cite{Jasmim2013} shows features characteristic of basaltic composition, therefore (27884) 1996 EZ1 could be basaltic as well. Additional spectra of this asteroid should be acquired to confirm its spectral type and to study potential variations in its spectrum.

\begin{figure*}[!h]
	\centering
	\begin{adjustbox}{clip,trim=0.cm 0cm 0.cm 0cm,max width=\textwidth}
		\input{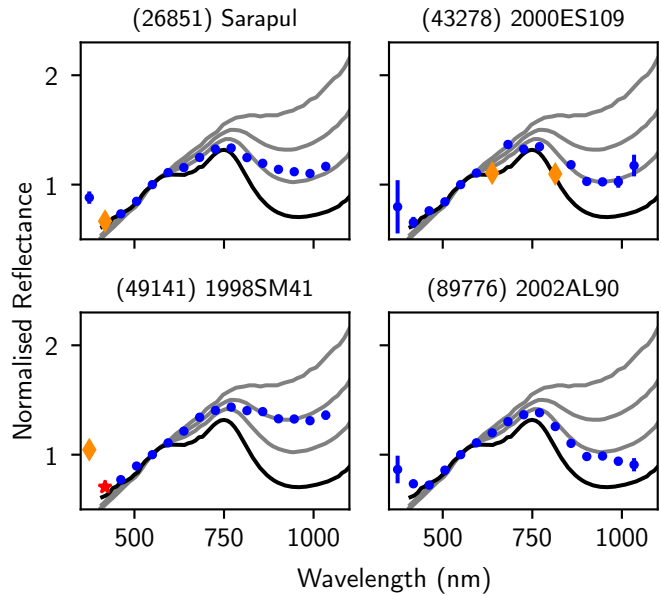}
	\end{adjustbox}
	\caption{Spectra of the four asteroids found as potential matches of the low and medium space-weathered EC~002 in \cite{galinier2023}, that are likely not V-types and that have a Gaia DR3 spectrum associated to an average S/N$\geq$30. The spectra are normalised at 550 nm. Black continuous line: spectrum of the powder sample of the meteorite, grey lines: spectra of the space-weathered modeled spectra of EC~002. The 16 bands of the Gaia asteroid spectra are given a colour and a symbol according to the value of the flag associated to the band: blue circle if flag=0, orange diamond if flag=1 and red star if flag=2.}
	\label{fig:pot_match}
\end{figure*}

As done in \cite{galinier2023}, we calculated the slope of the reflectance spectrum between 468.6 and 748~nm (gri slope), and the depth of the 1 $\mu$m band expressed as $\mathrm{R_z-R_i=R(\lambda = 893.2~nm)-R(\lambda = 748.0~nm)}$, for every Gaia DR3 asteroid and for the meteorite spectra. According to the position of the observed asteroids in the $R_z-R_i$ vs gri slope in Fig.\ref{fig:slope_zi}, it is likely that most of the potential matches determined in \cite{galinier2023} and that plot at $R_z-R_i\leq-0.3$ and gri slope $\leq$30 are V-types, given the concentration of observed V-types in this region. 
If we consider that these objects are likely V-types, we eliminate 94 objects, which leaves us with only 48 potential analogues of EC~002. Regarding the signal-to-noise ratio (S/N) of the Gaia DR3 spectrum \citep{galluccio2022} of these remaining 48 potential analogues, 37 have a S/N$\leq$21, which is below the acceptable threshold for a reliable classification of Gaia DR3 data \citep{Galinier2024}. Most were found potential analogues of EC~002 using a reduced $\chi^2$ as curve-matching parameter that favoured low-S/N asteroids to be found as matches \citep{galinier2023}. Acording to \cite{Andrae2010}, noise in the data makes the value of the reduced $\chi^2$ uncertain, thus these objects cannot be considered good potential analogues of the meteorite. Other six asteroids have a DR3 spectrum with an average S/N well below 30, and they should be considered with caution as well: (8243) Devonburr, (15623) Maitaneam, (18143) 2000 OK48, (33947) 2000 ML1, (53417) 1999 NP38, and (108139) 2001 GL11; with respective S/N of 22.6, 21.58, 24.9, 23.48, 21.95, and 21.74. If we consider that all these objects are likely not analogues of EC~002, we are left with only four potential analogues of EC~002 that we did not already observe: (26851) Sarapul, (43278) 2000~ES109 and (89776) 2002~AL90, found as potential matches of the low space-weathered meteorite; and (49141) 1998~SM41, found as potential match of the medium space-weathered meteorite. The spectra of these objects, along with those of the space-weathered meteorite spectra, are displayed in Fig.\ref{fig:pot_match}. These four objects could be observed in the NIR to determine their similarity with EC~002, but observing the other potential analogues determined in \cite{galinier2023} would likely lead to a majority of non-matching objects.

Secondly, the space-weathering model applied to EC~002 to search for potentially similar spectra among the asteroid population might not be adapted to andesitic meteorites. Indeed, applying the Hapke model as done by \cite{barrat2021} to simulate the space weathering of andesitic bodies is justified by the fact that andesites are dominated by silicates with grains larger than the wavelength \citep{galinier2023}. However, the Hapke model has been developed from the study of lunar samples \citep{Hapke2001}, and as space weathering affects materials differently depending on their composition \citep[e.g.][]{Lantz2017,brunetto2015,Zhang2022}, it is not guaranteed that the Hapke model perfectly simulates the space-weathering of andesitic material. In fact, from the study of regolith breccias of the H4-chondrite Fayetteville and the howardite Kapoeta, and the analysis of the space-weathering of S-type asteroids (243) Ida and (433) Eros, \cite{Gaffey2011} even states that lunar-type space weathering can explain the link between OCs and the S(IV)-subset of S-type asteroids, but is however rarely observed on asteroid surfaces. If the Hapke space-weathering model is not adapted to EC~002, this has an impact on the potential analogues found by comparison with the meteorite's modelled space-weathered spectra. It could therefore be interesting to perform space-weathering experiments on EC~002-like simulants, to fully understand how space weathering affects andesitic materials, and to better search for potential analogues of such meteorite in the asteroid population.

In the end, the objects showing the BIC and BAR parameters the closest to those of EC~002 are asteroids (10537) 1991~RY16 and (14390) 1990~QP10. These asteroids share spectral characteristics with EC~002, such as two broad bands centred around 1 and 2 $\mu$m and a band around 650 nm, clearly distinguishable for (10537) 1991~RY16. The band still has to be confirmed in the spectrum of (14390) 1990~QP10, since it is not clearly visible in the spectrum of \cite{Matlovic2020}, and its Gaia DR3 spectrum is affected by the`fake band' issue at 650 nm (see Fig.\ref{fig:lit_weird}).

The pronounced bands at 1 and 2 $\mu$m of asteroid (10537) 1991~RY16 lead \cite{Moskovitz2010} to suggest that the effect of space weathering on this asteroid might be negligible, as what is seen on (4) Vesta \citep[see e.g.][]{McCord1970,Pieters2012}. This hypothesis still has to be demonstrated, as space weathering affect the spectra of vestoids \citep{Fulvio2012,Marchi2010}. It is however a possibility, and if it is correct, then taking into account space weathering to try linking such object to meteorites with curve matching methods might not be necessary. This could explain why the small feature around 650 nm is visible in the spectrum of this asteroid, and does not get masked like that of EC~002 when Hapke's lunar-type space weathering model is applied to its powder spectrum (see Fig.\ref{fig:pot_match}).

Another possibility would be that the minerals at the origin of the 650 nm feature are somehow resistant to space weathering (the possible origin of this feature is discussed in the following). This is however purely speculative, as to date, there is not record in the literature of space weathering experiment performed on meteorites presenting such feature. As discussed for EC~002, space weathering could also impact (10537) 1991~RY16 and asteroids of similar compositions in a non-lunar way, possibly preserving the feature at 650 nm.

On the other hand, if the minerals responsible for the 650 nm feature are affected by space weathering, this could explain why so little asteroids appear to show this feature: it could be masked in the spectra. Therefore, the effect of space weathering on objects showing such a spectral feature should be further studied, and could help conclude on the link between (10537) 1991~RY16, (14390) 1990~QP10, and EC~002. Given our current knowledge and according to this study, which relies on spectral characteristics not affected by space weathering \citep{Gaffey2011}, these asteroids are the best analogues found for EC~002.

According to its spectrum and its position in the BIC–BAR parameter space, (7472) Kumakiri appears closer to an O-type asteroid than to EC~002, due to the near absence of the 2000 nm absorption band, consistent with its classification by \cite{Hicks2014}. According to \cite{barrat2021}, the modeled spectrum of a mixture composed of 80\% EC~002 powder and 20\% olivine provides a good match to that of O-type asteroids (see their supplementary material). In addition, EC~002 contains variable proportions of olivine xenocrysts depending on the sample considered \citep{barrat2021,Jin2024}. Since the powder sample used in \cite{barrat2021} and in the present study is devoid of olivine xenocrysts \citep{barrat2021}, (7472) Kumakiri and O-type asteroids may have compositions similar to more olivine-rich samples of EC~002. Furthermore, the two visible spectra of this asteroid reported by \cite{Duffard2009} show the feature around 650 nm. Therefore, (7472) Kumakiri cannot be entirely ruled out as a potential match to EC~002.

The presence of an absorption band at 650 nm can have various causes \citep{moskovitz2008_10537}. First, this band may be due to iron charge transfers between $Fe^{3+}$ and $Fe^{2+}$ in the pyroxene structure of an object \citep{moskovitz2008_10537}. Second, the presence of pyroxenes containing large amounts of chromium creates significant absorption features around 630 nm, which probably explains the presence of the band in the spectrum of the asteroid (10537) 1991~RY16 according to \cite{cloutis2018}. Third, when chromium is present in spinel (chromite), it can exhibit absorption features due to $Cr^{3+}$ electronic transitions. However, the transitions of $Fe^{2+}$ and $Cr^{3+}$ often lead to the appearance of other absorption bands around 550, 590, 690, or 2000 nm \citep{Cloutis2004}, which are not observed here. Finally, a combination of all these effects could produce such a band in the spectrum of (10537) 1991~RY16 and, if confirmed, in the one of (14390) 1990~QP10. A band around 650 nm is also observable in the spectrum of EC~002. Its origin is not discussed in \cite{barrat2021}, but the meteorite's pyroxenes are quite rich in Cr-bearing species (see their Table~S2), which can lead to the apparition of such absorption feature \citep[e.g.][]{moskovitz2008_10537,cloutis2018,cloutis2002}. Therefore, the small band around 650 nm of EC~002 could be explained by the presence of chromium-bearing pyroxene. We note that this feature alone is not diagnostic of an andesitic composition, since it is present in the spectrum of the meteorite LEW 86220 that belongs to the acapulcoite–lodranite group \citep{Lucas2019}.

\cite{cloutis2018} found that asteroid (10537) 1991~RY16 is the closest match to the ungrouped achondrite North-West Africa (NWA) 7325, based on the study of the spectral characteristics of the two bodies. It is however not a completely satisfactory match, since the absorption bands in the asteroid's spectrum are more consistent with low-calcium pyroxene, and NWA~7325 shows high-calcium pyroxene contents \citep{cloutis2018}. Knowing that EC~002 contains low-Ca pyroxene \citep{barrat2021}, we can conclude that (10537) 1991~RY16 is a better match to EC~002 than to NWA 7325.

Finally, \cite{leith2017} hypothesize that (10537) 1991~RY16, (7472) Kumakiri, and (14390) 1990 QP10 are fragments of differentiated planetesimals that have been destroyed or ejected from the Solar System. They are found to show more compositional diversity than (4) Vesta family members, suggesting that they may have formed at a different time or location in the protoplanetary disk. These interpretations and the position of all three objects in the BIC vs BAR space leads us to think that there could be a relationship between the meteorite EC~002 and the asteroids (10537) 1991~RY16, (14390) 1990~QP10, and (7472) Kumakiri.

These three objects are all located in the outer main belt. (10537) 1991~RY16 is in the so-called `pristine zone' \citep{Broz2013} close to the 5:2 mean motion resonance (MMR) with Jupiter. The `pristine zone' is a region between 2.82 and 2.96 au, which shows a number density of asteroids lower than the average of the main belt \citep{Tsirvoulis2018}. For this reason, \cite{Broz2013} suggested that the primordial distribution of asteroids could be best reflected in this region, and indeed \cite{Tsirvoulis2018} showed that the magnitude distribution of the primordial population of main belt asteroids with 9<H<12 is qualitatively well represented in the pristine zone's background population. However, this region was found to be contaminated by asteroids crossing resonances and drifting from adjacent regions \citep{Tsirvoulis2018}, which makes it not so `pristine'. Asteroid (10537) 1991~RY16 could therefore be primordial, or could have been implanted in this region via dynamical processes \citep{moskovitz2008_10537,RaymondIzidoro2017,Tsirvoulis2018,Avdellidou2022}. The other two asteroids have semi-major axis of about 3 and 3.3 au: (7472) Kumakiri is located close to the 7:3 MMR, and (14390) 1990~QP10 is close to the 2:1 MMR. If these objects are fragments of differentiated planetesimals, they could have been implanted in the main belt long after the formation and disruption of their parent body \citep{Avdellidou2022,Avdellidou2024}. The fact that these objects are all located in the outer main belt, along with the discovery of the first potential olivine-rich family in the `pristine zone' \cite{Galinier2024}, might suggest that new evidence of differentiation should be searched for in the outer main belt. The study of the rocks' strength and of the effectiveness of the resonances in sending material to Earth-crossing orbits is beyond the scope of this work.

Considering these three asteroids as potential matches to EC~002, we still identify very few objects with spectral characteristics similar to those of this meteorite. Several explanations may account for this result. It is possible that most planetesimals bearing andesitic crusts have disappeared from the main belt, because of erosion, collisional destruction, or accretion to growing planets; as suggested by \cite{barrat2021} and \cite{moskovitz2008_10537} for EC~002 and (10537) 1991~RY16 respectively. Another possibility is that andesitic asteroids are still present in the main belt but have not yet been identified as such. For example, to explain the lack of mantle-like asteroids in the main belt, \cite{Schultz2025} explored the mixing of mantle and carbonaceous material. They found that olivine and pyroxene-rich asteroids might be present in much more significant amount than previously thought, but are not identified as such because of spectral mixing. The same phenomenon could be happening with andesitic bodies, modifying or hiding their characteristic spectral features. Space weathering may also affect the surfaces of these bodies, but as discussed above, dedicated experiments are required to conclude on this matter.

Finally, the apparent scarcity of asteroids displaying andesitic features may simply reflect the fact that such objects have not been actively searched for to date. Indeed, (10537) 1991~RY16, (7472) Kumakiri, and (14390) 1990 appear to have been discovered by chance during observational campaigns primarily targeting basaltic V-type asteroids \citep{moskovitz2008_10537,Duffard2009,Hicks2014,leith2017,Hardersen2018,Migliorini2018,Matlovic2020}. It is possible that other objects presenting similar spectral characteristic as these bodies exist in the main belt. In fact, we can see on Fig.\ref{fig:slope_zi} that (10537) 1991~RY16, (7472) Kumakiri and (14390) 1990~QP10 all plot in the same area of the $R_z-R_i$ vs gri slope space. The fact that these asteroids do not plot outside the main asteroid cluster, and that they are not isolated in this region, may indicate that other asteroids in the main belt share similar spectral characteristics. Besides, since the effects of space weathering depend on the mineralogical composition of asteroids \citep{Marchi2010,Fulvio2012,Zhang2022}, and considering that these bodies share a similar composition, we may reasonably assume that they are affected by space weathering in a similar way. Therefore, it should be possible to search for spectral analogues of (10537) 1991~RY16, (7472) Kumakiri, and (14390) 1990~QP10 among the asteroid population without modelling space weathering.

The Gaia DR3 dataset should be used with caution to identify such objects, due to the previously mentioned issues affecting it such as the `fake band'. Indeed, if the presence of a feature near 650 nm is used as a criterion to identify these asteroids, the `fake band' issue could generate numerous false positives. Upcoming surveys, such as the Gaia Data Release 4 (DR4) and SPHEREx \citep{Bock2026_spherex}, will provide VIS and NIR spectra for thousands of asteroids. Combined with ground-based VIS and NIR observations, these data could enable the identification of asteroids sharing spectral characteristics with (10537) 1991~RY16, (7472) Kumakiri, and (14390) 1990~QP10, and help clarify their potential link with EC~002.

\section{Conclusions and perspectives}
\label{sec:ccl}

To try and find analogues of andesite EC~002 in the main belt asteroid population, we performed observations in the NIR with the IRTF of 20 potential analogues determined by \cite{galinier2023}. We classified their VISNIR spectra and obtained that 16 of these asteroids are spectroscopic V-types, asteroid (31060) 1996~TB6 is an A-type, and asteroid (18780) Kuncham belongs to the S-complex. The study of the diagnostic spectral features of these asteroids and their comparison with those of EC~002 showed that none of the observed asteroids appears to be a good match for the meteorite. In fact, based on the analysis of the observed asteroids and the comparison of their parameters with those of the other potential analogues, it appears that most of the potential analogues of EC~002 identified in \cite{galinier2023} likely do not correspond to the meteorite, most of them likely being basaltic objects. This can be explained by the fact that basaltic and andesitic objects show similar spectra in the VIS, apart from the weak band at 650 nm, only typical of the latter ones, and that can be confused with the `fake band' sometimes affecting the Gaia DR3 data. Moreover, the low S/N of some of the potential analogues of the meteorite selected in \cite{galinier2023} make their selection as potential analogue less reliable. Finally, the way space-weathering processes affect andesitic bodies should be further studied, to improve the search for analogues of EC~002.

The objects that appear to best match the spectrum of EC~002 are asteroids (10537) 1991~RY16 and (14390) 1990~QP10, both located in the outer main belt. Asteroid (7472) Kumakiri could match a more olivine-rich sample of EC~002, and cannot be completely ruled out as a match of the meteorite. Theories concerning the origin of these asteroids suggest that they could be fragments of differentiated planetesimals \citep{moskovitz2008_10537,leith2017}, similar to what has been proposed regarding the origin of EC~002 \citep{barrat2021}. Authors suggest that the planetesimals at the origin of these asteroids and that of EC~002 had likely disappeared from the main belt, due to their accretion to other bodies or to their destruction \citep{moskovitz2008_10537,barrat2021}. Similarities in the spectral features and in the origin and evolution theories of these asteroids and of EC~002 suggest a potential link between these objects. More asteroids spectrally similar to these bodies could exist in the main belt, and they could potentially form a population of andesitic bodies. This finding is consistent with the theory of \cite{Migliorini2018} suggesting that asteroid (14390) 1990~QP10 might be part of a new taxonomic class. Such objects could be searched for in the main belt, for example taking advantage of the coming Gaia Data Release 4 and SPHEREx surveys \citep{Bock2026_spherex}, and performing observations in the widest wavelength range possible.

\section*{Acknowledgements}

MG and MD acknowledge financial support from CNES and the Action Specifique Gaia.
MD is Leverhulme Visiting Professor at the University of Leicester with financial support from the Leverhulme Trust (UK).
MD, CA, and LG acknowledge financial support from the ANR ORIGINS (ANR-18-CE31-0014). MG, CA and MD were Visiting Astronomers at the Infrared Telescope Facility, which is operated by the University of Hawaii under contract 80HQTR19D0030 with the National Aeronautics and Space Administration.
This work has made use of data from the European Space Agency (ESA) mission
{\it Gaia} (\url{https://www.cosmos.esa.int/gaia}), processed by the {\it Gaia}
Data Processing and Analysis Consortium (DPAC,
\url{https://www.cosmos.esa.int/web/gaia/dpac/consortium}). Funding for the DPAC
has been provided by national institutions, in particular the institutions
participating in the {\it Gaia} Multilateral Agreement. 
This work is based on data provided by the Minor Planet Physical Properties Catalogue (MP3C) of the Observatoire de la Côte d'Azur.
The authors thank the two anonymous reviewers for their comments that helped significantly improve this article.



\appendix

\section{Observations of potential analogues of EC~002}
\label{app:obs}

\begin{center}
	\begin{sideways}
		\begin{minipage}{0.9\textheight} 
			\centering
			\captionof{table}{Table of observations of potential analogues of EC~002. The table includes object number, the match type (space weathering or not), the proper orbital elements of the asteroids (semi-major axis, eccentricity and inclination), the geometric albedo, the magnitude in the V band, the date of the observations (dd-mm-yy), the solar analogues used, and the airmass. The spectral type corresponds to the classification of the observed asteroids in the Bus-DeMeo scheme with the MIT online classifier \citep{demeo2009}. The two observed spectra of (24286) 1999 XU188 are designated as (a) and (b). \label{tab:obs}}
			\setlength{\tabcolsep}{5pt}
			\begin{tabular}{l c c c c c c c c c c}
				\hline
				Asteroid & Match type & $a_p$ (au) & $e_p$ & $sin(i_p)$ (°) & pV & Vmag & Date (UT) & Solar Analogues & Airmass & Spectral type  \\ 
				\hline 
				1643  & no SW &           2.4892 & 0.1379 & 0.0809 & 0.164	      &  17.4      &      28-01-24   & SA102-1081 & 1.100 & Vw \\
				1946   & no SW &        2.2936 & 0.1902 & 0.1304 & 0.362	&	   16.3       &28-01-24 & SA102-1081 & 1.111  & Vw \\		   
				3188  & no SW &       2.2894 & 0.0919 & 0.0844 & 0.425	&	16.0    &     29-07-23 &  HD215017 & 1.430 & V \\		   
				3869  & no SW &          2.4524 & 0.1008 & 0.0920 & 0.144		  &    17.9      &08-03-23 & SA107684, HD 112049  & 1.800 & V \\		   
				4302  & no SW &         2.4565 & 0.1428 & 0.0941 & 0.240	&	 17.0   &       08-12-23 & SA115-271, HD6470 & 1.100  & V \\		   
				6003 & low SW &       2.3444 & 0.0935 & 0.1116 & -	&	  17.4    &    08-03-23 & SA107684, HD137272 & 1.300 & V \\	   
				6789 & low SW &         2.3380 & 0.1392 & 0.1021 & -		  &      16.8      & 08-12-23 &  SA112-1333, HD198259 & 1.251   & Vw \\	   
				8827 & no SW &        2.3142 & 0.1047 & 0.0866 & -		    &  16.9  &       08-12-23 & SA112-1333, HD198259 & 1.260   & Vw \\	   
				10671  & no SW &       2.4351 & 0.1387 & 0.0832 & 0.272	&	  16.8    &    18-12-23 & SA98-978, HD259516 & 1.029 & Vw \\		   
				12551  & no SW &      2.4230 & 0.0970 & 0.0955 & -	&	16.5	   &   02-02-23 & HD76752 & 1.220 & U \\		   
				14511 & low SW &        2.3833 & 0.0779 & 0.1325 & 0.286	&	 17.0       &      03-10-23 & SA115-271 & 1.286 & V \\	   
				15989 & no SW &         2.3262 & 0.1417 & 0.0871 & 0.609	&	  17.0      & 28-01-24 &  SA102-1081 & 1.185 & Vw \\		   
				17056 & no SW &      2.2142 & 0.0556 & 0.0498 & 0.340	&	17.4     &     03-11-23 & SA93101 & 1.073 & V \\		   
				18780  & low SW &       2.2293 & 0.0813 & 0.0617 & -	&	  17.1     & 03-10-23 & SA115-271 & 1.407 & S-complex, S \\	   
				20454  & no SW &        2.4707 & 0.1191 & 0.1126 & 0.257   	&	  17.6    &      03-11-23 & SA93101 & 1.068  & V \\	   
				24286  (a) & no SW &     2.2992 & 0.1975 & 0.1021 & - &	 17.2      & 29-07-23 & SA112-1333, HD215017 & 1.420  & V \\		   
				24286  (b)  & no SW &    2.2992 & 0.1975 & 0.1021 & -	&	     17.2     &     03-10-23 & SA112-133 & 1.291 & V \\	   
				27884 & low SW &      2.3643 & 0.0901 & 0.1224 & 0.390	&	17.3     &         03-11-23 & SA93101 & 1.084  & U \\
				30426 & low SW &    2.4184 & 0.1436 & 0.1030 & 0.301	&	     17.6       &         08-12-23 & SA112-1333, HD211706 & 1.080  & V \\	   
				31060  & medium SW &   2.7283 & 0.3004 & 0.1878 & 0.380	&	 15.5    &            03-11-23 & SA93101 & 1.119 & A \\	   
				45787  & low SW &	  2.4325 & 0.1002 & 0.1112 & 0.613 &		  17.6          &             08-12-23 &  SA115-271, HD6470 & 1.160  & V \\			
				\hline						
			\end{tabular}
			
			\begin{tablenotes}
				\small
				\item[\bfseries Note:] The geometric albedo values are the best values reported in MP3C. SW stands for space weathering. The w in spectral type Vw stands for weathered, it flags asteroids with slopes greater than 0.25 \citep{demeo2009}. U stands for Undetermined. The spectrum of (31060) 1996~TB6 has already been published and classified in \cite{Delbo2026}, but it is reported here for the sake of completeness, as it is considered a potential match for EC~002.
			\end{tablenotes}
			
		\end{minipage}
	\end{sideways}
\end{center}

\begin{figure}[p]
	\centering
	\begin{adjustbox}{clip,trim=1cm 2.1cm 1cm 2.5cm,max width=\textwidth}
		\input{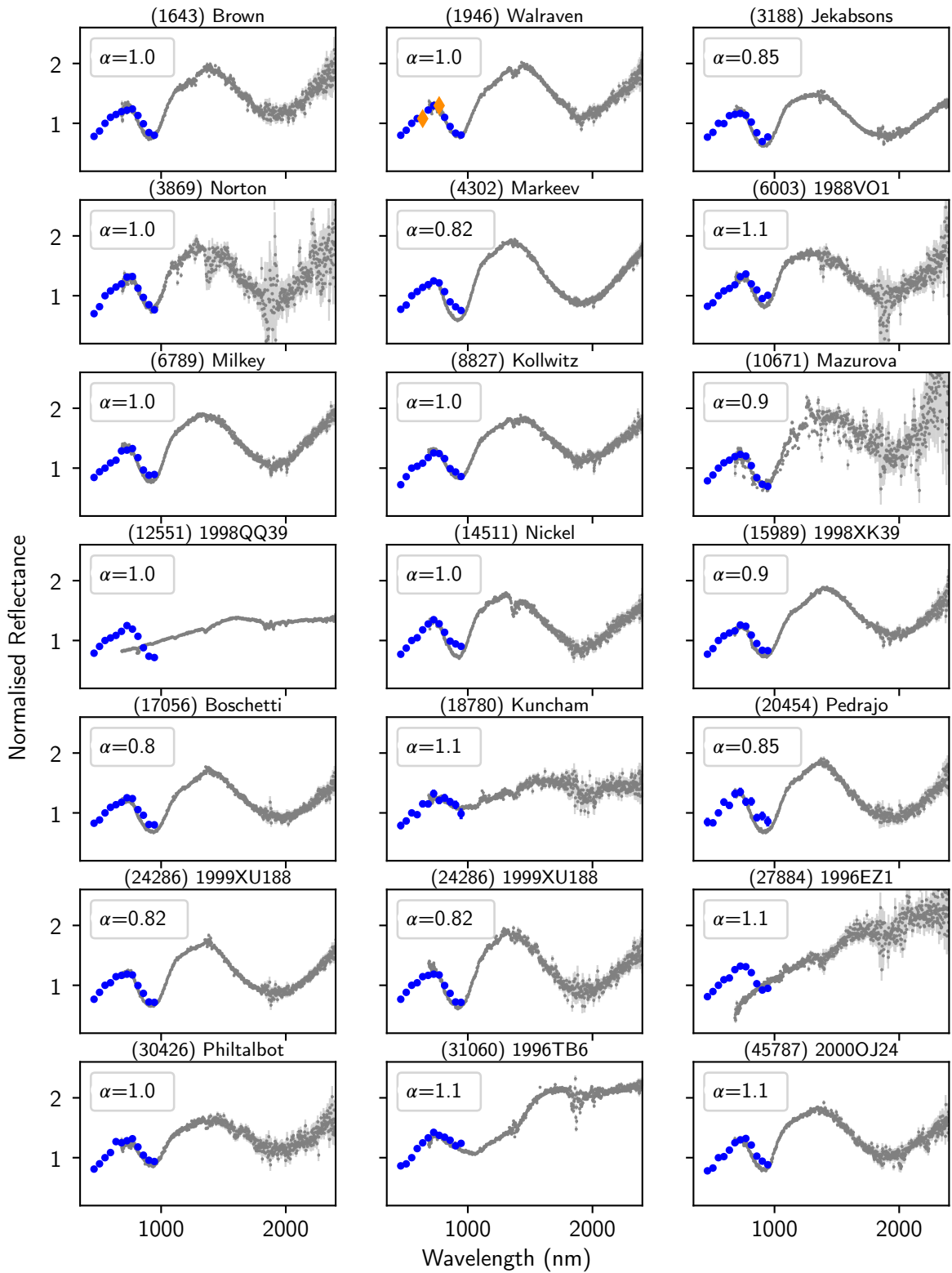}
	\end{adjustbox}
	\caption{NIR (grey dots, uncertainties given as grey shaded area) and Gaia DR3 spectra considered in the range [462, 946] nm of the observed asteroids. The multiplicative factor applied to each NIR spectrum to align it with the Gaia DR3 spectrum is given in the legend of each sub-figure, for each asteroid. The bands of Gaia DR3 spectra are assigned a colour and symbol according to their flag value: blue circle if flag=0 and orange diamond if flag=1.}
	\label{fig:EC002_analogues_nir}
\end{figure}

\section{Scaling NIR spectra with Gaia DR3 VIS spectra: automatic procedure}
\label{app:visir_merging}

The Gaia DR3 dataset is known to be affected by several issues, including a systematic reddening that affects the spectra in the RP region, longward of 700 nm \citep{galluccio2022,galinier2023,Balossi2023MasterThesis,Galinier2024PhD}. This reddening must be taken into account when scaling NIR spectra with the VIS Gaia DR3 spectra, particularly when classifying asteroids. On the contrary, the BP part of the DR3 spectra was found not to be affected by such reddening \citep{Galinier2024PhD}. As a test, we performed a scaling of our observed NIR spectra with the VIS Gaia DR3 spectra using an automated technique \citep{Christou2021,popescu2019}. First, the NIR spectra were normalised at 1000 nm as described in Sect.\ref{sec:class}. Then, we resampled the VIS and NIR spectra in their overlapping region between 695 and 950 nm using a cubic smoothing spline (python function \textit{csaps}). We then used the function
\begin{equation}
		\chi^2 = \sum_{i}^{N} (R_i^{VIS}-\alpha \times R_i^{NIR})^2
	\label{E:chi2}
\end{equation}
to scale the NIR part with the visible part of the spectrum, with $R_i^{VIS}$ and $R_i^{NIR}$ respectively the VIS and NIR reflectance measurements at the i-th wavelength position in the overlapping region, and $\alpha$ varying between 0.85 and 1.05, following \cite{Christou2021}. For each spectrum, the automatically selected $\alpha$ factor corresponds to the minimum $\chi^2$ value. The $\alpha$ factors obtained with this procedure and with the visual scaling method described in Sect.\ref{sec:class} are given in Table.\ref{tab:alpha_auto} for comparison.

First, we note that the automated scaling method systematically yields an $\alpha$ factor larger than the one estimated visually. The difference between the two factors exceeds 0.7 for eight asteroids, including 12551 and 27884, which show a discrepancy between their VIS and NIR spectra and therefore cannot be properly scaled. We classified the other six asteroids as V-types, as described in Sect.\ref{sec:class}.

\begin{figure}[!htb]
	\centering
	\begin{subfigure}{0.49\textwidth}
		\centering
		\begin{adjustbox}{clip,trim=0.5cm 0.cm 0.5cm 0.5cm,max width=\linewidth}
			\includegraphics{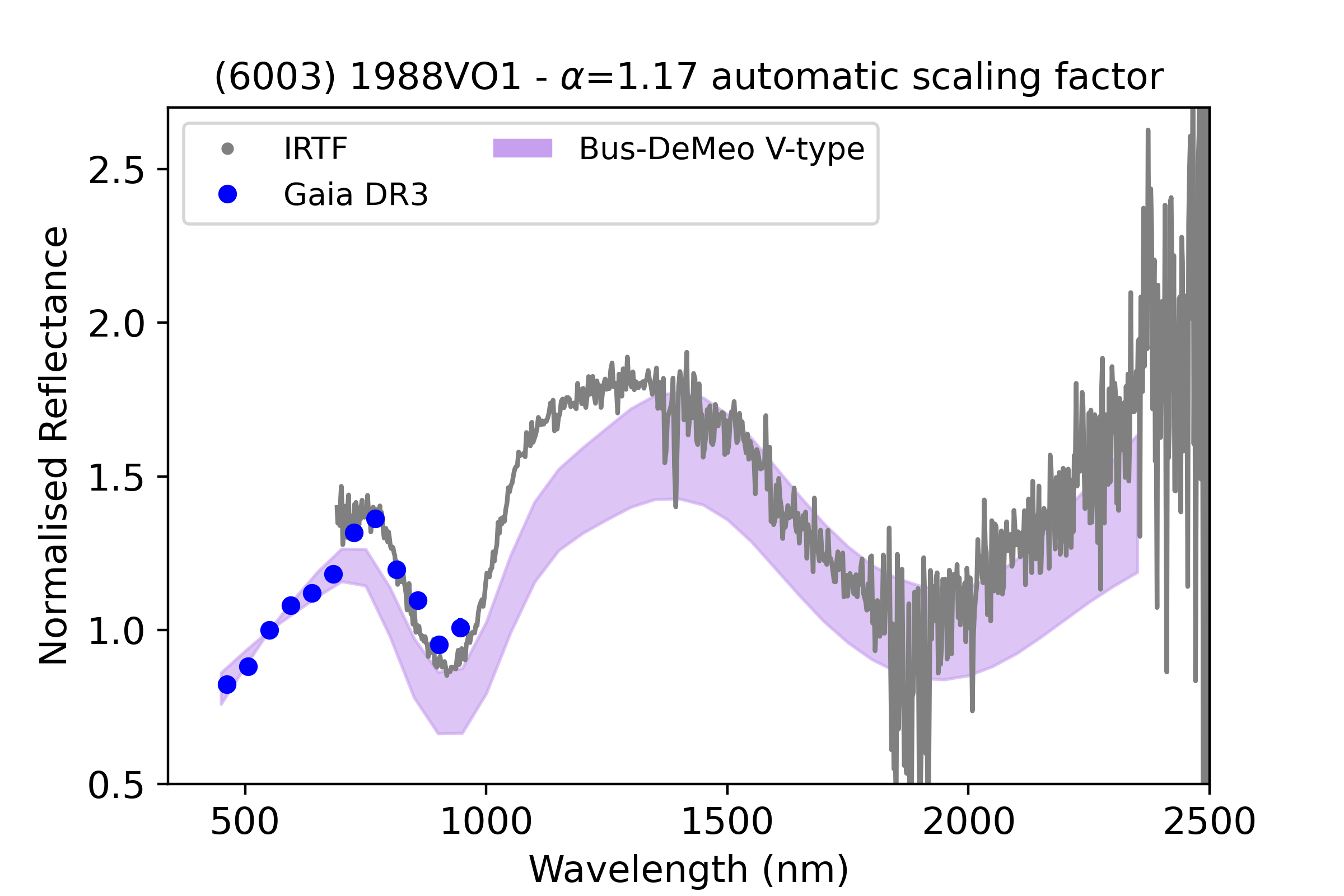}
		\end{adjustbox}
	\end{subfigure}
	\hfill
	\begin{subfigure}{0.49\textwidth}
		\centering
		\begin{adjustbox}{clip,trim=0.5cm 0.cm 0.5cm 0.5cm,max width=\linewidth}
			\includegraphics{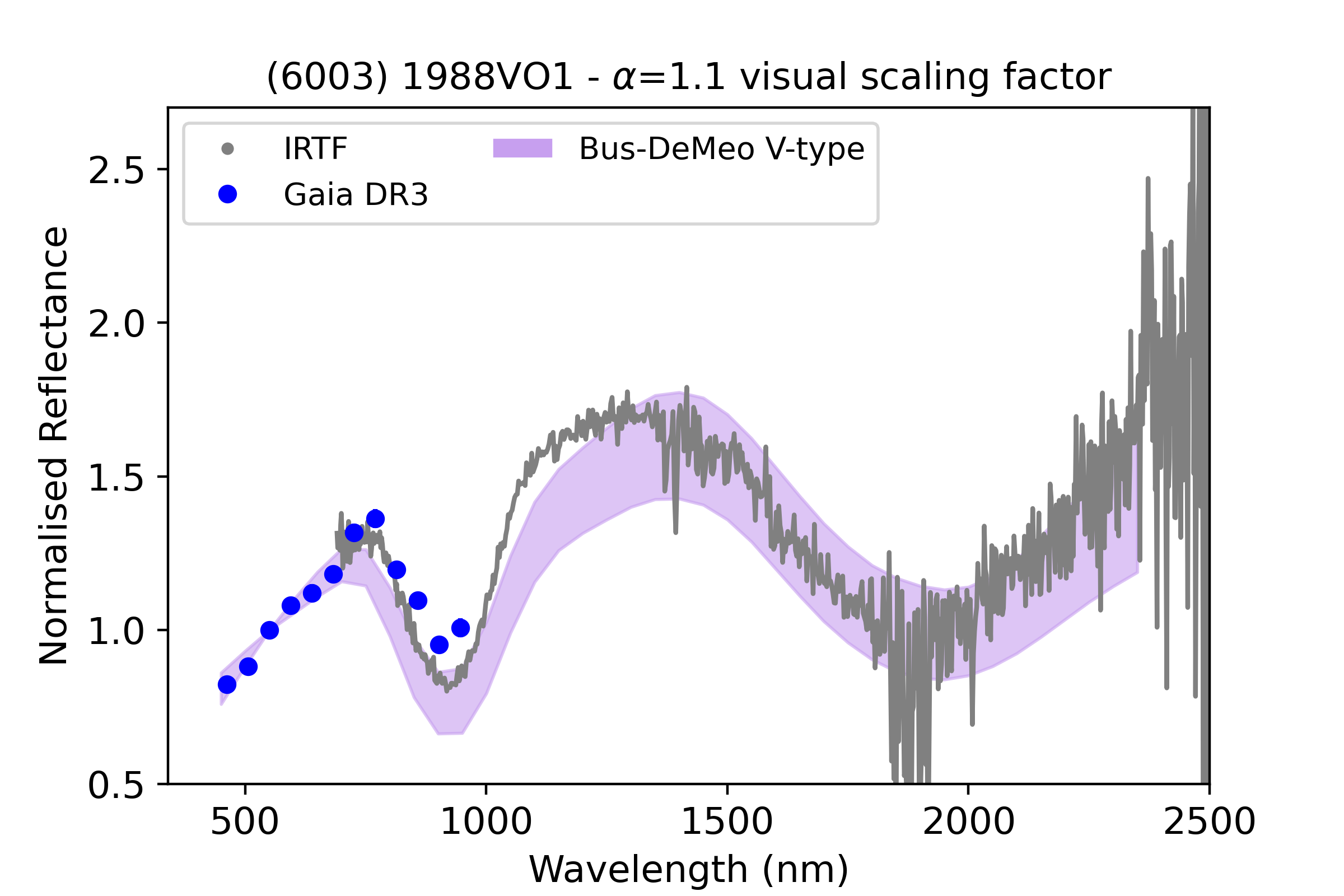}
		\end{adjustbox}
	\end{subfigure}
	\caption{Observed NIR spectra of asteroid (6003) 1998~VO1 (grey dots), and Gaia DR3 spectrum considering the range from 462 to 946 nm (blue dots). Left panel: VISNIR merging using an automated scaling procedure; right panel: VISNIR merging applying a factor $\alpha$ visually determined. The V-type template spectrum of the Bus-DeMeo taxonomic scheme \citep{demeo2009} is shown as a purple shaded area.}
	\label{fig:visnir_merging_comp}
\end{figure}

In Fig.\ref{fig:visnir_merging_comp} is shown the full VISNIR spectrum of asteroid (6003) 1998~VO1, after the automatic (left panel) and the visual (right panel) scaling. The V-type template spectrum of the Bus-DeMeo taxonomic scheme \citep{demeo2009} is given as a reference. We can see that the Gaia DR3 spectrum of (6003) 1998~VO1 follows the V-type template well in the BP range, and shows a `jump' in reflectance values around 700 nm. This is characteristic of the reddening that can affect DR3 spectra in the RP region, longward 650 nm \citep{galluccio2022,Galinier2024PhD}. Since the automated scaling method relies on the RP region to scale the NIR spectrum to the VIS, this results in the production of a full VISNIR spectrum artificially shifted towards higher reflectance values. For asteroid (6003) 1998~VO1, scaling the NIR part with an $\alpha$ factor determined visually allows for a better scaling between the NIR and the BP part of the DR3 spectrum, even if it does not completely correct for the reddening of the VIS spectrum.


Scaling automatically a NIR spectrum with a reddened VIS spectrum could affect its classification, as the produced VISNIR spectrum would appear reddened. This could be an issue mostly for spectral types on which the slope plays an important role in the classification, i.e featureless spectra, and it should not affect us much here. However, classifying the VISNIR spectrum of asteroid (31060) 1996~TB6 automatically scaled with the MIT tool gives the following result: `visual inspection needed: either D-type or A-type'. From visual inspection and according to \cite{Delbo2026}, (31060) 1996~TB6 is an A-type. The MIT tool gives this inconclusive classification because the automatically scaled VISNIR spectrum of (31060) 1996~TB6 shows a redder slope than that of A-types, comparable to that of D-types. Therefore, using a visually determined $\alpha$ parameter to scale the spectra helps mitigate the effect of the Gaia DR3 reddening issue and improves the classification of asteroids. As a side note, the procedure we use to compare asteroid spectra with that of EC~002 and to determine band centres and band area ratios should be only mildly affected by the scaling, since we remove the spectral continuum which is the component most impacted by this issue.

\begin{table}[!h]
	\centering
	\caption{Scaling factor $\alpha$ obtained using the automated and the visual scaling procedure for each observed asteroid. (a) and (b) refer to the two observed spectra of the asteroid (24286) 1999 XU188. \label{tab:alpha_auto}}
	\begin{tabular}{l c c}
		\hline
		Observed asteroid & Automatic scaling $\alpha$ & Visual scaling $\alpha$  \\ 
		\hline 
		(1643) Brown  &  1.03  &  1.00 \\
		(1946) Walraven  &  1.02  &  1.00 \\
		(3188) Jekabsons  &  0.87 &  0.85 \\
		(3869) Norton  &  1.03 &  1.00 \\
		(4302) Markeev  &  0.90  &  0.82 \\
		(6003) 1988 VO1  &  1.17  &  1.10 \\
		(6789) Milkey  &  1.03 &  1.00 \\
		(8827) Kollwitz  &  1.01 &  1.00 \\
		(10671) Mazurova  &  0.92 &  0.90 \\
		(12551) 1998 QQ39  &  1.11  &  1.00 \\
		(14511) Nickel  &  1.07 &  1.00 \\
		(15989) Anusha  &  0.95 &  0.90 \\
		(17056) Boschetti  &  0.87 &  0.80 \\
		(18780) Kuncham  &  1.12 &  1.10 \\
		(20454) Pedrajo  &  0.92 &  0.85 \\
		(24286) 1999 XU188 (a)  & 0.84 &  0.82 \\
		(24286) 1999 XU188 (b)  & 0.83 &  0.82 \\
		(27884) 1996 EZ1  &  1.42 &  1.10 \\
		(30426) Philtalbot  &  1.06 &  1.00 \\
		(31060) 1996 TB6  &  1.14 &  1.10 \\
		(45787) 2000 OJ24  &  1.18 &  1.10 \\
		\hline
	\end{tabular}
\end{table}

\newpage

\section{V-types: vestoids or not?}
\label{app:V}

\cite{Oszkiewicz2023} found that remnants of multiple differentiated planetesimals may be present in the inner main belt, not only (4) Vesta. These basaltic asteroids are expected to exhibit spectral differences from (4) Vesta, reflecting variations in pyroxene chemistry as observed in (1459) Magnya \citep{Hardersen2004,Ieva2016,Mansour2020,Oszkiewicz2023}. This asteroid is believed to have originated from a different parent body from (4) Vesta \citep{Hardersen2004}, and \cite{Oszkiewicz2023} identified a number of inner main-belt objects showing spectral characteristics, such as the band depth, suggesting that they may not belong to the Vesta family.

As a preliminary analysis, we compared the NIR spectra of our observed asteroids to those of (4) Vesta and (1459) Magnya of \cite{demeo2009}. To do so, we resampled the spectra between 690 and 2400 nm using steps of 5 nm, and we normalised the resampled spectra at 1000 nm. Then, we calculated the following $\chi^2$ as defined by \cite{Nedelcu2007}:
\begin{equation}
 	\mathrm{\chi_{red}^{2} = \frac{1}{N}  \sum_{i}^{N} \frac{(R^{Vesta}_{i}-R_{i})^{2}}{R_{i}}},
 	\label{eq:chi2_V}
\end{equation}
with N the number of wavelengths, R the NIR reflectance spectrum of our observed asteroid, and $R^{Vesta}$ the spectrum of Vesta (or Magnya) to which is compared the asteroid. Two $\chi^2$ are then calculated, one from the comparison of the asteroid with Vesta and one from its comparison with Magnya. The smallest value of the $\chi^2$ gives the best match of each asteroid between (4) Vesta and (1459) Magnya. We did not consider the VIS spectra of the asteroids in this calculation, as their Gaia DR3 spectra might be impacted by above-mentioned issues such as a reddening or a `fake band', which could impact the $\chi^2$. Besides, NIR spectra are enough to perform mineralogical characterisations of asteroids and to compare (1459) Magnya with (4) Vesta, as done by \cite{Hardersen2004}.

\begin{figure}[p]
	\centering
	\begin{adjustbox}{clip,trim=1cm 2.1cm 1cm 2.5cm,max width=\textwidth}
		\input{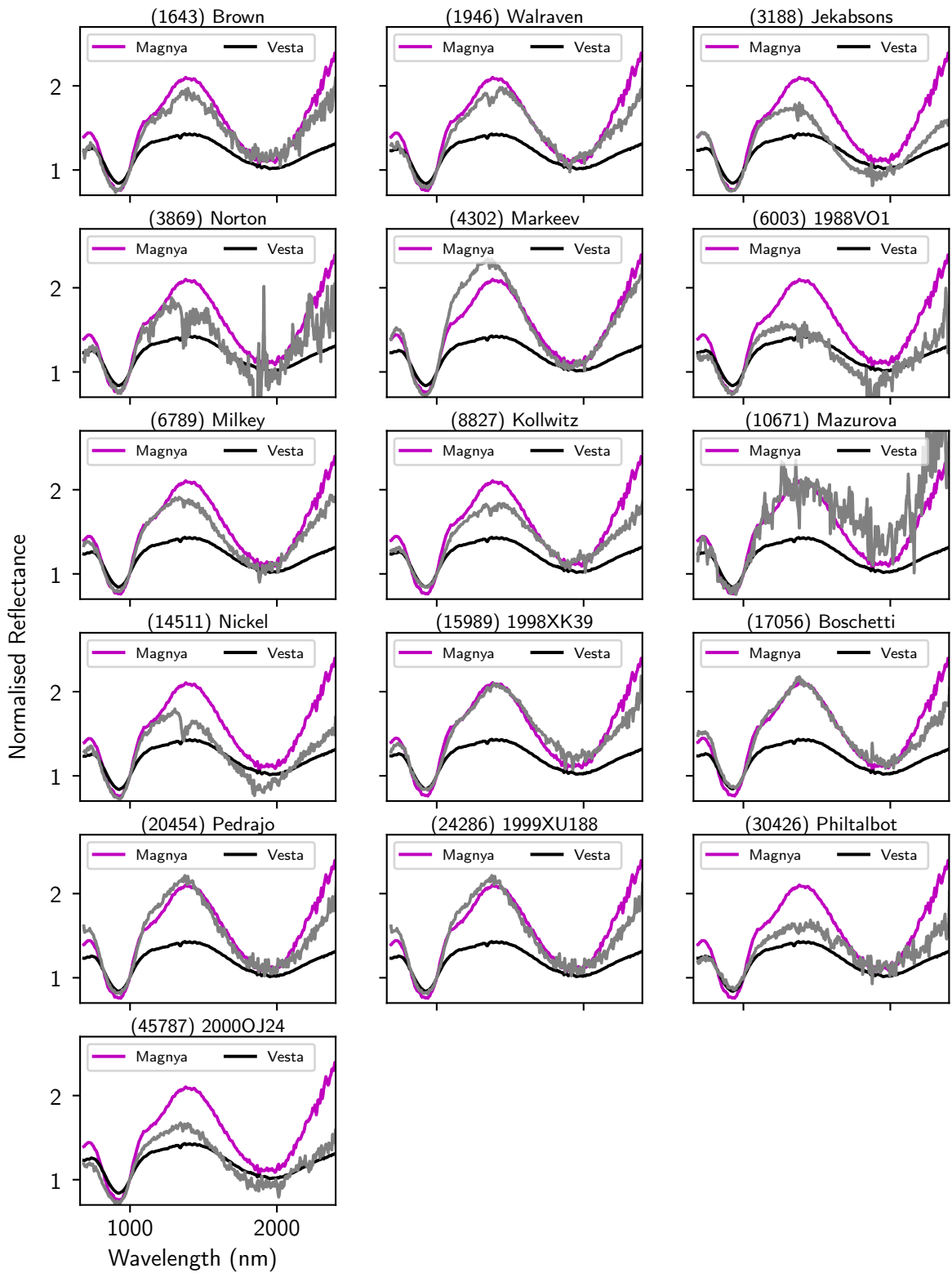}
	\end{adjustbox}
	\caption{NIR normalised and resampled spectra of the observed asteroids classified V-types (grey), compared with the spectra of (4) Vesta (black) and (1459) Magnya (magenta) of \cite{demeo2009}.}
	\label{fig:vesta_magnya_V}
\end{figure}

According to the $\chi^2$ calculation, we find that asteroids (1643) Brown, (1946) Walraven, (4302) Markeev, (6789) Milkey, (8827) Kollwitz, (10671) Mazurova, (15989) Anusha, (17056) Boschetti, (20454) Pedrajo and (24286) 1999 XU188 show a better spectral correspondence with (1459) Magnya than with (4) Vesta, as can be seen on Fig.\ref{fig:vesta_magnya_V}. According to this preliminary analysis, (20454) Pedrajo might therefore be an interloper in the Vesta family. These asteroids could be fragments of a differentiated planetesimal distinct from (4) Vesta, which is consistent with what \cite{Oszkiewicz2023} suggested. A deeper analysis has to be performed to conclude on their belonging or not to the Vesta family, but this study is outside the scope of this work.

\newpage

\printcredits

\bibliographystyle{cas-model2-names}

\bibliography{newbib}



\end{document}